\documentclass[journal=jctcce,manuscript=article]{achemso}

\usepackage{amsmath}
\usepackage{amssymb}
\usepackage{amsfonts}
\usepackage{bm}
\usepackage{bbm}
\usepackage{braket}
\usepackage{mathtools}
\usepackage[Symbol]{upgreek}
\usepackage{hyperref}
\hypersetup{
    colorlinks=true,
    linkcolor=blue,
    citecolor=red,
    filecolor=magenta,      
    urlcolor=cyan
}

\DeclareMathOperator{\pr}{\mathrm{p}^{+}}

\title{Symmetry-Projected Nuclear-Electronic Hartree--Fock: Eliminating Rotational Energy Contamination}

\setkeys{acs}{ maxauthors = 0 }

\author{Robin Feldmann}
\affiliation{ETH Z\"urich, Department of Chemistry and Applied Biosciences, Vladimir-Prelog-Weg 2, 8093 Z\"urich, Switzerland}
\author{Alberto Baiardi}
\affiliation{ETH Z\"urich, Department of Chemistry and Applied Biosciences, Vladimir-Prelog-Weg 2, 8093 Z\"urich, Switzerland}
\author{Markus Reiher}
\email{mreiher@ethz.ch}
\affiliation{ETH Z\"urich, Department of Chemistry and Applied Biosciences, Vladimir-Prelog-Weg 2, 8093 Z\"urich, Switzerland}

\begin{document}
\begin{center}
\date{{\bf This work is dedicated to\\ Professor Roland Lindh on the occasion of his 65th birthday}\\ August 31, 2023}
\end{center}

\begin{abstract}
We present a symmetry projection technique for enforcing rotational and parity symmetries in nuclear-electronic Hartree--Fock wave functions,
which treat electrons and nuclei on equal footing.
The molecular Hamiltonian obeys rotational and parity-inversion symmetries, which are, however, broken by expanding in Gaussian basis sets that are fixed in space.
We generate a trial wave function with the correct symmetry properties by projecting the wave function onto representations of the three-dimensional rotation group, i.e., the special orthogonal group in three dimensions SO(3).  As a consequence, the wave function becomes an eigenfunction of the angular momentum operator which (i) eliminates the contamination of the ground state wave function by highly excited rotational states arising from the broken rotational symmetry, and (ii) enables the targeting of specific rotational states of the molecule.
We demonstrate the efficiency of the symmetry projection technique by calculating energies of the low-lying rotational states of the H$_2$ and H$_3^+$ molecules.
\end{abstract}


\maketitle

\section{Introduction}

Solving the molecular Schr\"odinger equation by considering both nuclei and electrons on equal footing\cite{Hammes-Schiffer2020_Review} incorporates effects beyond the Born--Oppenheimer approximation in molecular simulations.\cite{Born-Oppenheimer1927} 
Those effects are the result of the coupling between nuclear and electronic degrees of freedom, i.e., nonadiabatic and nuclear quantum effects. Hence, phenomena which are governed by these effects benefit from such pre-Born-Oppenheimer methods. 
  
Formally, pre-Born-Oppenheimer theories harbor the advantage over Born--Oppenheimer-based methods that the nonrelativistic interaction potential between nuclei and electrons is given exactly by pairwise Coulomb interactions. By contrast, in the case of the latter methods, the potential entering the vibrational Hamiltonian is an $N$-body potential that must be approximated with fitting or interpolation algorithms\cite{Bowman2009_Permutationally,Carrington2020_NeuralNetworkPES}. This also implies that the coupling between nuclear degrees of freedom of different kinds (e.g., rovibronic coupling) is treated directly in pre-Born--Oppenheimer theories.   
Moreover, also the nuclear-permutation and spin symmetries are considered from the outset.

Both the nuclei and the electrons can be described with single-particle functions, which are referred to as orbitals.
Thomas pioneered this approach in 1969~\cite{Thomas1969_protonic1,Thomas1969_protonic2,Thomas1970_protonic3,Thomas1971_protonic4} and Petitt proposed the first nuclear electronic Hartree--Fock method~\cite{Pettitt1986_FirstMcHF} which was developed into a practical method by Nakai and coworkers~\cite{Nakai1998_NOMO-Original}.
Since then, many electronic-structure methods were extended to treat nuclei quantum mechanically:
including M{\o}ller--Plesset perturbation theory~\cite{Swalina2004-PreBO_MP2,nakai2007nuclear,Pavosevic_OOMP2-Multicomponent,Fajen2021_MulticomponentMP4}, configuration interaction~\cite{Valeev2004_ENMO,Hammes-Schiffer2002,Patrick_2015_EN-MFCI,Brorsen2020_SelectedCI-PreBO,Brorsen2020_multicomponentCASSCF}, density matrix renormalization group \cite{Muolo2020_Nuclear,Feldmann2022_QuantumProton}, coupled cluster \cite{Nakai2003,Ellis2016_Development,Pavosevic2019_MulticomponentCC,Pavosevic2021_MulticomponentUCC,Hammes-Schiffer2021_MulticomponenDF,Brorsen2022_Triples}, and density functional theory \cite{Hammes-Schiffer2012_MulticomponentDFT,Hammes-Schiffer2017_MulticomponentDFT,Brorsen2018_TransferableDFT,HammesSchiffer2019_GGA-DFT}.

The molecular Hamiltonian possesses continuous symmetries.
Translational and rotational symmetries must therefore be respected by any wave function ansatz.
Solving the molecular Schr\"odinger equation based on atom-centered single-particle functions, however, breaks the continuous symmetries of the Hamiltonian.
A consequence of this symmetry breaking is that variational methods yield inaccurate energies even for very flexible and fully optimized wave functions \cite{Nakai2005_EliminationRotTrans,Muolo2020_Nuclear}.
Although the translational energy can be subtracted exactly from a non-relativistic Hamiltonian, breaking the symmetry can introduce spurious excited states in its spectrum\cite{Valeev2004_ENMO}.

Nakai et al.\  showed that the problem of rotational energy contamination can be remedied by approximately eliminating the rotational energy from the Hamiltonian~\cite{Nakai2005_EliminationRotTrans,Nakai2006_Elimination2,Sutcliffe2005_CommentNakai}.
In their approach, the rigid-body rotational energy is expanded around an equilibrium geometry in a Taylor series which was shown to converge below 1~mHa already at zeroth order \cite{Nakai2006_Elimination2}. 
However, this approach is limited solely to rotational ground states.
Moreover, it effectively changes the Hamiltonian from the exact molecular Hamiltonian to an approximate one in which all nuclei are distinguishable.

Heller and Blanco introduced the projection onto rotational states within the Born--Oppenheimer approximation \cite{Heller1983_Projection}.
Later, we extended this approach by projecting pre-Born--Oppenheimer wave functions expressed in terms of explicitly correlated Gaussians onto states exhibiting proper rotational symmetry \cite{muolo2019}.

In this work, we introduce a method to eliminate exactly the rotational energy contamination to the nuclear-electronic Hartree--Fock (NE-HF) wave function by enforcing the correct rotational symmetry of the wave function through symmetry projection. The projection is carried out in the projection-after-variation fashion, i.e., the orbitals are optimized by standard nuclear-electronic Hartree--Fock calculations and the wave function is subsequently symmetry projected.
This method represents a first step toward accurately calculating vibrational and rotational spectra of small to moderately sized molecules without the Born--Oppenheimer approximation.
We
adopt the approach from nuclear structure theory\cite{ring_2004nuclear,Sheikh_2000symmetry,sheikh2021_SymmetryReview,Bally2021_Projection} and present its theoretical foundations in
Sec.~\ref{sec:theory}. Then, Sec.~\ref{sec:implementation} provides the necessary details for its implementation.
After a description of computational details in Sec.~\ref{sec:ComputationalDetails}, we apply our approach to calculate the energies of rotational states of H$_2$ and H$^+_3$ in Sec.~\ref{sec:results}.
We selected H$_2$ and H$^+_3$ for scrutinizing our approach since the symmetry properties of these molecules are well understood and highly accurate reference data are available in the literature.

\section{Theory}
\label{sec:theory}

\subsection{Symmetry properties of the molecular Schr\"odinger equation}
\label{subsec:FullMolecularSchrodingerEquation}

The translation-free molecular Hamiltonian for a system comprising $N_\mathrm{p}$ particles with masses $m_i$, charges $q_i$, and positions $\mathbf{r}_i\in\mathbb{R}^3$, is given as~\cite{Muolo2020_Nuclear}
\begin{equation}
  \hat{H}(\mathbf{r}) =
   - \sum_{i}^{N_{\text{p}}} \left(\frac{1}{2 m_i}-\frac{1}{2M}\right) \bm{\nabla}^2_i
   + \sum_{i<j}^{N_{\text{p}}} \frac{q_i q_j}{|\mathbf{r}_i -\mathbf{r}_j|} 
   + \sum_{i<j}^{N_{\text{p}}} \frac{1}{M} \bm{\nabla}_i\bm{\nabla}_j,
  \label{eq:TranslationallyInvariantHam}
\end{equation}
where $\mathbf{r}\in \mathbb{R}^{3N_\mathrm{p}}$ collects the positions of all particles, $M = \sum_i^{N_\text{p}} m_i$ is the total mass, and $\bm{\nabla}_{i}=(\nabla_{i x},\nabla_{i y},\nabla_{i z})^T$ is the derivative operator with respect to the position of particle $i$.
While the Born--Oppenheimer electronic Hamiltonian treats nuclear coordinates as fixed parameters, the molecular Hamiltonian treats the nuclear positions explicitly as quantum mechanical operators.
As a result, the molecular Hamiltonian is of higher symmetry than the Born--Oppenheimer electronic Hamiltonian.
Specifically, the Born--Oppenheimer Hamiltonian obeys a discrete point group symmetry that originates from the fixed nuclear positions, whereas the molecular Hamiltonian has continuous translational and rotational symmetry, as well as discrete spatial-inversion symmetry.
Translational symmetry specifies that the system remains unchanged under a constant shift, $\mathbf{a}$, 
\begin{equation}
  \hat{H}(\mathbf{r}) = \hat{H}(\mathbf{r}+\mathbf{a}) \quad \text{with} \quad 
    \mathbf{a} \in \mathbb{R}^{3N_\mathrm{p}} \, .
  \label{eq:TranslationalSymmetry}
\end{equation}
Eq.~(\ref{eq:TranslationallyInvariantHam}) fulfills Eq.~(\ref{eq:TranslationalSymmetry}) because the derivative operator is invariant upon overall translations and the Coulomb interaction depends only on the relative positions of the particles, which remain unchanged upon translation.
Rotational symmetry demands that the system remains unchanged under a rotation,
\begin{equation}
  \hat{H}(\mathbf{R}\mathbf{r}_1, \mathbf{R}\mathbf{r}_2,\dots) = \hat{H}(\mathbf{r}_1, \mathbf{r}_2, \dots) 
    \quad \text{with} \quad \mathbf{R}\ \in \mathrm{SO}(3).
  \label{eq:RotationalSymmetry}
\end{equation}
Here, $\mathrm{SO}(3)$ is the group of orthogonal three-by-three matrices with determinant one, defined as
\begin{equation}
  \mathrm{SO}(3) = \{  \mathbf{R} \in \mathbb{R}^{3\times3} : 
              \mathbf{R} \mathbf{R}^\mathrm{T} = \mathbbm{1}, \mathrm{det}(\mathbf{R})=1
          \} \, .
\label{eq:DefSO3}
\end{equation}
Both the derivative operator and the Coulomb interaction are isotropic and, therefore, they are invariant upon symmetry transformations of the $\mathrm{SO}(3)$ group.
The group of spatial inversions, $C_I=\{\mathbbm{1},\mathbf{I}\}$, consists of only two elements: the identity $\mathbbm{1}$ and the inversion matrix $\mathbf{I}$, which is defined as $\mathbf{I}\mathbf{r} = -\mathbf{r}$.
It is easy to prove that the Hamiltonian is invariant also upon spatial inversion, that is,
\begin{equation}
  \hat{H}(-\mathbf{r}) = \hat{H}(\mathbf{r}).
  \label{eq:InversionHamiltonian}
\end{equation}
The symmetries of the Hamiltonian dictate that the wave function must transform according to irreducible representations of the corresponding symmetry groups.
As a result, exact solutions to the Schr\"odinger equation can be labeled according to these irreducible representations, which serve as good quantum numbers.
The ground-state translational energy has already been removed in the  Hamiltonian of Eq.~(\ref{eq:TranslationallyInvariantHam}), and therefore, we omit the corresponding symmetry label.
The irreducible representations of the $\mathrm{SO}(3)$ group are labeled by the angular momentum quantum number $J=0, 1, \dots$ and its projection onto the $z$-axis, denoted by $M_J = -J, -J+1, \dots, J$.
Additionally, we denote the parity quantum number as $p=\pm 1$ for even and odd parity, respectively.

An additional wave function symmetry is given by the spin, which is represented by the
$\mathrm{SU}(2)$ group with the spin quantum numbers for particles of type $I$ given as $S_I=0,\frac{1}{2},1\dots$, and $M_{S,I}=-S_I, -S_I+1,\dots,S_I$. 
In our approach, we rely on the unrestricted nuclear-electronic Hartree--Fock ansatz, which only allows us to specify $M_{S,I}$, as Slater determinants (or permanents) are generally not eigenfunctions of the $\hat{S}^2_I$ operator.
The wave function ansatz reads
\begin{equation}
  \Psi^\mathrm{HF}_{M_{S,1},\dots,M_{S,N_t}}(\mathbf{r}_{1},\dots,\mathbf{r}_{N_\text{t}}) = \prod_I^{N_{\text{t}}} \Phi_{I,M_{S,I}} (\mathbf{r}_{I}) ~,
  \label{eq:ansatz}
\end{equation}
with the number of particle types, $N_\text{t}$, and $\mathbf{r}_I$, the vector that contains the coordinates of all particles of type $I$. 
$\Phi_{I,M_{S,I}}$ is constructed as a properly symmetrized product of nuclear or electronic molecular orbitals, $\phi_{Isi}$,
\begin{equation}
  \Phi_{I,M_{S,I}}(\mathbf{r}_I) = \frac{1}{\sqrt{N_{I}!}} \mathcal{S}_\pm \left( \prod^{s_I}_s \prod_i^{N_{Is}} \phi_{Iis}(\mathbf{r}_{Iis}) \right),
  \label{eq:SingleTypeAnsatz}
\end{equation}
where $s_I$ is the single-particle spin quantum number, while $\mathbf{r}_{Iis}$ labels the
coordinate of the $i$-th particle of type $I$ with spin-projection $s$.
$N_{Is}$ denotes the number of particles of type $I$ with spin-projection on the $z$-axis $s$.
The index $s$ runs over all possible values of the magnetic quantum number $m_{s,I}$.
We note that $M_{S,I}$ is fixed by the specified number of particles of type $I$ for each spin $s$.
The (anti)symmetrization operator, $\mathcal{S}_\pm$, enforces the correct permutational symmetry of the product of orbitals, i.e., by antisymmetrization for fermions and by symmetrization for bosons. 

The molecular orbitals of particle type $I$ are constructed as a linear combination of $L_I$ Gaussian-type orbitals (GTOs), $G_{I\mu}$, with pre-optimized Gaussian widths, contraction coefficients, and shifts as
\begin{equation}
  \phi_{Iis}(\mathbf{r}_{Iis}) = \sum_\mu^{L_I} G_{I\mu}(\mathbf{r}_{Iis}) C_{I\mu s i},
  \label{eq:AO2MO}
\end{equation}
where we restrict ourselves to solid-harmonic contracted GTOs.
The expansion into GTOs, however, comes with the severe drawback that the wave function ansatz breaks the rotational, translational, and parity symmetries of the Hamiltonian. That is, it does not transform according to irreducible representations of these symmetry groups.
In fact, the centers of the GTOs are located at fixed points in space -- usually at the position of the nuclei predicted based on a Born--Oppenheimer calculation.
Since Gaussian basis sets have anisotropic distributions when not centered at the center of mass of the molecule, they inherently break the rotational symmetry.

\subsection{Restoring wave function symmetries by symmetry projection}
\label{subsec:SymmetryProjection}

Since the use of GTOs breaks the rotational symmetry, the exact expansion of the Hartree--Fock wave function is given as a linear combination of eigenstates of the Hamiltonian with various $J$ states.
This symmetry-breaking results in unreasonably high energies \cite{Nakai2005_EliminationRotTrans}, even for fully variationally optimized wave functions, since states with arbitrarily high $J$ values can contribute significantly to the wave function.
We address this issue by introducing a method to project out all $J$ states in the Hartree--Fock wave function except the one of interest.
For simplicity, we focus on the rotational symmetry.

We denote the Hilbert space in which the exact solution to the molecular Schr\"odinger equation is defined as $\mathcal{H}$.
A complete basis for $\mathcal{H}$ can be constructed from the eigenstates $|\Psi_{n,J,M_J}\rangle$ of $\hat{H}$, with $n$ representing all quantum numbers except the rotational ones such that
\begin{equation}
  \mathcal{H} = \mathrm{span}\{\ket{\Psi_{n,J,M_J}}\}_{n,J=0,1,\dots,\infty,M_J=-J,\dots,J}.
  \label{eq:CompleteBasisEigenfunctions}
\end{equation}
As a result, $\mathcal{H}$ decomposes into a series of invariant subspaces associated with the irreducible representations of $\mathrm{SO}(3)$ and the remaining quantum numbers $n$ as
\begin{equation}
  \mathcal{H} = \bigoplus_{n=0}^\infty \bigoplus_{J=0}^\infty \mathcal{H}^{nJ},
  \label{eq:HilbertSpaceDecomposition}
\end{equation}
where the subspaces, $\mathcal{H}^{nJ}$, are at least $2J+1$ dimensional, depending on other symmetries (contained in $n$) or accidental degeneracies.

The Hartree--Fock wave function, $|\Psi^\mathrm{HF}\rangle$, can be expanded in the basis $\{|\Psi_{n,J,M_J}\rangle\}$ as
\begin{equation}
  \vert \Psi^\mathrm{HF}\rangle = \sum_n^\infty\sum_{J}^{\infty} \sum_{M_J=-J}^Jd_{n,J,M_J} \vert \Psi_{n,J,M_J} \rangle \, .
  \label{eq:SymmBrokenHfExpansion}
\end{equation}
We consider Eq.~(\ref{eq:SymmBrokenHfExpansion}) which is easily generalized to a general symmetry-broken wave function as a linear combination of functions that transform according to irreducible representations of the rotational symmetry group.
A projection operator, $\hat{P}^J_{M_J,M_J}$, can remove the contributions from all states that are not associated with a specific set of $J$ and $M_J$ values.
The resulting wave function, $\hat{P}^J_{M_J,M_J}|\Psi^\mathrm{HF}\rangle$, becomes an eigenfunction of the total angular momentum operator $\hat{J}^2$ with quantum numbers $J$ and $M_J$.
However, depending on the orientation of the symmetry broken state $|\Psi^\mathrm{HF}\rangle$, the composition of Eq.~(\ref{eq:SymmBrokenHfExpansion}) varies and it may be that the projection operator projects onto the null vector if $d_{n,J,M_J}=0$ for all $n$.
To overcome this issue, we consider the projection on all possible $M_J$ for a given $J$.
In Dirac's bra-ket notation, the projection operator can be written as
\begin{equation}
  \hat{P}^J_{K,L} = \sum_n^\infty |\Psi_{n,J,K}\rangle\langle\Psi_{n,J,L}|
  \label{eq:ProjectionOperator}
\end{equation}
from which the following key properties follow
\begin{align}
    \hat{P}^J_{M,N}\hat{P}^I_{K,L} &= \delta_{IJ}\delta_{NK} \hat{P}^J_{M,L} \, , 
    \label{eq:P_property1}
    \\
    (\hat{P}^J_{M,L})^\dagger      &= \hat{P}^J_{L,M} \, .
    \label{eq:P_property2}
\end{align}
$\hat{P}^J_{KL}$ is an orthogonal projection operator in the strict mathematical sense, i.e., it satisfies the properties $\hat{P}^2=\hat{P}=\hat{P}^\dagger$, only for $K=L$.
When $K\neq L$, the operator is sometimes referred to as the ``shift operator'' or ``transfer operator''~\cite{Bally2021_Projection}.
We now apply the operator to the Hartree--Fock wave function in Eq.~(\ref{eq:SymmBrokenHfExpansion}) and obtain
\begin{equation}
  \hat{P}^J_{K,L} |\Psi^\mathrm{HF}\rangle = \sum_n^\infty d_{n,J,L} |\Psi_{n,J,K}\rangle.
  \label{eq:ProjHF}
\end{equation}
To remove the dependence on the orientation of $|\Psi^\mathrm{HF}\rangle$ and the specified $M_J$ value in the ansatz, we need to diagonalize the Hamiltonian in the subspace spanned by the projected wave functions $\hat{P}^J_{K,L} |\Psi^\mathrm{HF}\rangle$ for all $K$ and $L$.
To that end, we first note that the Hamiltonian and overlap matrices expressed in the $\hat{P}^J_{K,L} |\Psi^\mathrm{HF}\rangle$ basis are block-diagonal, with one block per $J$ value \cite{Bally2021_Projection}:
\begin{align}
  \langle\Psi^\mathrm{HF}|  \hat{P}^{J\dagger}_{K,L} \hat{H} \hat{P}^J_{M,N}| \Psi^\mathrm{HF}\rangle &= \delta_{KM}  \langle\Psi^\mathrm{HF}| \hat{H} \hat{P}^J_{L,N}| \Psi^\mathrm{HF}\rangle \, , \\
  \langle\Psi^\mathrm{HF}|  \hat{P}^{J\dagger}_{K,L} \hat{P}^J_{M,N}| \Psi^\mathrm{HF}\rangle         &= \delta_{KM}  \langle\Psi^\mathrm{HF}| \hat{P}^J_{L,N}| \Psi^\mathrm{HF}\rangle \, ,
\end{align}
where we employed $[\hat{P}^J_{K,L},\hat{H}]=0$ and the properties in Eqs.~(\ref{eq:P_property1}) and (\ref{eq:P_property2}).
The above equations provide a way to define the Hamiltonian and overlap matrices within a given angular momentum irreducible representation of dimension $2J+1$, defined as follows
\begin{align}
  H^J_{KL} &= \langle\Psi^\mathrm{HF}| \hat{H} \hat{P}^J_{K,L}| \Psi^\mathrm{HF}\rangle, \label{eq:HWH} \\
  W^J_{KL} &= \langle\Psi^\mathrm{HF}| \hat{P}^J_{K,L}| \Psi^\mathrm{HF}\rangle. \label{eq:HWW}
\end{align}
Diagonalization of the Hamiltonian in this subspace yields the projected Hartree-Fock wave function as
\begin{equation}
   |\Psi^\mathrm{PHF}_{J,M_J}\rangle = \sum_{K=-J}^J f_K \hat{P}^J_{M_J,K}|\Psi^\mathrm{HF}\rangle,
   \label{eq:ProjectedHF}
\end{equation}
which is an eigenfunction of $\hat{J}^2$ and is independent on the orientation of the symmetry-broken state $|\Psi^\mathrm{HF}\rangle$.
The coefficients $f_K$ can be determined by solving the Hill--Wheeler equations~\cite{ring_2004nuclear,Bally2021_Projection}
\begin{equation}
  \sum_L H^J_{KL} f_L = E_{J} \sum_L W^J_{KL} f_L \, ,
  \label{eq:HillWheeler}
\end{equation}
and the energies are the generalized eigenvalues, $E_J$.
By comparing Eq.~(\ref{eq:SymmBrokenHfExpansion}) and Eq.~(\ref{eq:ProjHF}) we note that via the symmetry projection, we can eliminate the contribution of higher-lying rotational states.
For this reason, the symmetry projected energy is a better variational estimate of the exact one.
We now proceed by deriving the explicit expression for the projection operator.

\subsection{Angular momentum and parity projection operators}
\label{subsec:AngularMomentumParity}

A rotation in three-dimensional space can be uniquely represented by three consecutive rotations around the Euler angles, denoted as $\Omega = (\alpha, \beta, \gamma)$.
In this work, we adopt the convention of active rotations\cite{Gregory1987_GuideToRotations} where the three rotations are performed in the following order: (1) rotate the system by an angle $0 \leq \gamma \leq 2\pi$ about the $z$-axis, (2) rotate the system by an angle $0 \leq \beta \leq \pi$ about the $y$-axis, and (3) rotate the system by an angle $0 \leq \alpha \leq 2\pi$ about the $z$-axis.
With this convention, we can write down a generic rotation $\mathbf{R}\in \mathrm{SO}(3)$ as a product of rotation matrices according to\cite{Gregory1987_GuideToRotations}
\begin{equation}
  \mathbf{R}(\Omega) = \mathbf{R}_z(\alpha) \mathbf{R}_y(\beta) \mathbf{R}_z(\gamma)\,,
  \label{eq:Rotation}
\end{equation}
where the rotation matrices read
\begin{equation}
  \mathbf{R}_y(\phi) = 
    \begin{pmatrix}
      \cos \phi & 0 & \sin \phi \\
          0     & 1 &      0    \\
     -\sin \phi & 0 & \cos \phi \\
    \end{pmatrix} \, ,
\end{equation}
and
\begin{equation}
  \mathbf{R}_z(\phi) = 
    \begin{pmatrix}
      \cos \phi  & -\sin \phi & 0 \\
      \sin \phi  &  \cos \phi & 0 \\
      0          &      0     & 1 \\
    \end{pmatrix} \, .
\end{equation}
We introduce the generator of rotations around an axis $i \in \{x,y,z\}$, denoted by $\mathbf{J}_i$, as the derivative of the rotation matrices, $\mathbf{R}_i\in \mathrm{SO}(3)$, according to
\begin{equation}
  \mathbf{J}_i = - \mathrm{i} \frac{\mathrm{d} }{\mathrm{d} \phi} \mathbf{R}_i(\phi) \Bigg{|}_{\phi=0}\,.
  \label{eq:AngularMomentumRotation}
\end{equation}
The $\mathbf{J}_i$ operators are $3 \times 3$ matrix representations of the well-known angular-momentum operators (in Hartree atomic units).
Any three-dimensional rotation matrix can be parametrized in terms of the generators as
\begin{equation}
  \mathbf{R}(\Omega) = \mathrm{e}^{-\mathrm{i}\alpha \mathbf{J}_z}
                       \mathrm{e}^{-\mathrm{i}\beta \mathbf{J}_y}
                       \mathrm{e}^{-\mathrm{i}\gamma \mathbf{J}_z} \, .
  \label{eq:GeneratorParametrization}
\end{equation}
Since the action of $\mathbf{J}_i$ is only defined on $\mathbb{R}^3$, we introduce the representations of the generators of rotation for a generic Hilbert space as $\hat{J}_i$ and we write the general representation of the rotation operator as
\begin{equation}
  \hat{R}(\Omega) = \mathrm{e}^{-\mathrm{i}\alpha \hat{J}_z}\mathrm{e}^{-\mathrm{i}\beta \hat{J}_y}   \mathrm{e}^{-\mathrm{i}\gamma \hat{J}_z},
  \label{eq:GenericRotation}
\end{equation}
where $\hat{R}(\Omega)$ and $\hat{J}_i$ act on many-particle and single-particle Hilbert spaces.
The action of the rotation operator of Eq.~(\ref{eq:GenericRotation}) on an arbitrary angular momentum eigenstate $\ket{J,M_J}$ can be expressed with the closure of the angular momentum basis as
\begin{equation}
    \hat{R}(\Omega) \ket{J,M_J}= \sum_{K=-J}^J\ \langle J,K | \hat{R}(\Omega) | J, M_J \rangle  \ket{J,K} = \sum_{K=-J}^J D^J_{K,M_J}(\Omega) \ket{J,K},
\end{equation}
where the expansion coefficients are given by the Wigner D-matrices, $D^J_{K,M_J}(\Omega)$~\cite{Gregory1987_GuideToRotations}.
With this, we can derive the explicit expression of the action of the angular momentum projection operator onto the symmetry-broken Hartree--Fock wave function. To that end, we write down the action of the rotation operator from Eq.~(\ref{eq:GenericRotation}) on the symmetry-broken Hartree--Fock wave function from Eq.~(\ref{eq:SymmBrokenHfExpansion}) which was expanded in eigenstates of the Hamiltonian as
\begin{equation}
    \hat{R}(\Omega)|\Psi^\mathrm{HF}\rangle = \sum_n^\infty\sum_{J}^{\infty} \sum_{M_J=-J}^J\sum_{K=-J}^Jd_{n,J,M_J} D^J_{K,M_J}(\Omega)|\Psi_{n,J,K}\rangle,
    \label{eq:rotatedHF}
\end{equation}
where we expanded a rotated angular momentum state as a sum of eigenstates with the same $J$.
We exploit the orthogonality of the Wigner D-matrices, i.e.,
\begin{equation}
  \int \mathrm{d}\Omega\ D^{I*}_{L,N}(\Omega)D^J_{K,M}(\Omega) = \frac{8 \pi^2}{2J+1} \delta_{IJ}\delta_{LK}\delta_{NM},
  \label{eq:OrthogonalityWigner}
\end{equation}
with
\begin{equation}
   \int\mathrm{d}\Omega = \int_0^{2\pi} \mathrm{d}\alpha
                          \int_0^{\pi}\mathrm{d}\beta \sin\beta
                          \int_0^{2\pi}\mathrm{d}\gamma \, .
  \label{eq:IntegrationSolidAngle}
\end{equation}
We multiply Eq.~(\ref{eq:rotatedHF}) by $D^{J*}_{L,N}(\Omega)$ and integrate over $\Omega$, which yields
\begin{equation}
  \int \mathrm{d}\Omega\ D^{J*}_{L,N}(\Omega)\hat{R}(\Omega)|\Psi^\mathrm{HF}\rangle 
    = \frac{8 \pi^2}{2J+1} \sum_n^\infty d_{n,J,N} |\Psi_{n,J,L}\rangle,
  \label{eq:intermediate1}
\end{equation}
after exploiting Eq.~(\ref{eq:OrthogonalityWigner}).
The right-hand side of Eq.~(\ref{eq:intermediate1}) is equivalent to Eq.~(\ref{eq:ProjHF}) up to the constant pre-factor.
Hence, we can identify the projection operator as
\begin{equation}
    \hat{P}^J_{KL} = \frac{2J+1}{8\pi^2}\int \mathrm{d}\Omega \ D^{J*}_{KL}(\Omega)\ \hat{R}(\Omega).
\end{equation}
The energy of the projected wave function can be calculated by solving the Hill--Wheeler equation, Eq.~(\ref{eq:HillWheeler}), for the Hamiltonian operator
\begin{equation}
  H^J_{KL} = \frac{2J+1}{8\pi^2}\int \mathrm{d}\Omega \ D^{J*}_{KL}(\Omega)\   
    \langle\Psi^\mathrm{HF}| \hat{H} \hat{R}(\Omega)| \Psi^\mathrm{HF}\rangle,
  \label{eq:HW_H_explicit}
\end{equation}
and the overlap
\begin{equation}
    W^J_{KL} = \frac{2J+1}{8\pi^2}\int \mathrm{d}\Omega \ D^{J*}_{KL}(\Omega)\   \langle\Psi^\mathrm{HF}| \hat{R}(\Omega)| \Psi^\mathrm{HF}\rangle,
    \label{eq:HW_S_explicit}
\end{equation}
where the integration in the matrix element and the overlap of the Hartree--Fock wave function is carried out over the electronic and nuclear coordinates.
To project onto states with a given parity, we define the parity projection operator as~\cite{ring_2004nuclear}
\begin{equation}
  \hat{P}_p = \frac{1}{2}(\hat{\mathbbm{1}} + p \hat{I}),
  \label{eq:ParityProjection}
\end{equation}
which can be applied to the Hartree--Fock wave function as
\begin{equation}
 \begin{aligned}
    |\Psi^\mathrm{PHF}_{J,M_J,p}\rangle &= \sum_K f_K \hat{P}^J_{K,M_J}\hat{P}_p |\Psi^\mathrm{HF}\rangle \\
    &=  \sum_K f_K  \frac{2J+1}{16\pi^2}\int \mathrm{d}\Omega \ D^{J*}_{KL}(\Omega)(\hat{R}(\Omega) + p \hat{I}\hat{R}(\Omega)) |\Psi^\mathrm{HF}\rangle.
 \end{aligned}
\end{equation}
Hence, the matrix elements to be evaluated are
\begin{align}
    H^{Jp}_{KL} &= \frac{2J+1}{16\pi^2}\int \mathrm{d}\Omega \ D^{J*}_{KL}(\Omega)\   \langle\Psi^\mathrm{HF}| \hat{H}\hat{R}(\Omega) + p \hat{H}\hat{I}\hat{R}(\Omega)| \Psi^\mathrm{HF}\rangle,\\
    W^{Jp}_{KL} &= \frac{2J+1}{16\pi^2}\int \mathrm{d}\Omega \ D^{J*}_{KL}(\Omega)\   \langle\Psi^\mathrm{HF}| \hat{R}(\Omega) + p \hat{I}\hat{R}(\Omega)| \Psi^\mathrm{HF}\rangle.
\end{align}
These previous are the key to our symmetry-projected NE-HF method.

\section{Implementation of the projected NE-HF method}
\label{sec:implementation}

\subsection{Discretization of the projection operator}
\label{subsec:Discretization}

We discretize the integrals entering the Hill--Wheeler equations, and since the parity projection is straightforward as opposed to the rotational projection, we disregard the former and write the projected matrix elements as
\begin{equation}
  \begin{aligned}
    H^J_{KL} &= \frac{2J+1}{8\pi^2} \sum_g w(\Omega_g) D^{J*}_{KL}(\Omega_g)\langle\Psi^\mathrm{HF}| \hat{H} \hat{R}(\Omega_g)| \Psi^\mathrm{HF}\rangle,\\
    W^J_{KL} &= \frac{2J+1}{8\pi^2}\sum_g w(\Omega_g) D^{J*}_{KL}(\Omega_g)  \langle\Psi^\mathrm{HF}| \hat{R}(\Omega_g)| \Psi^\mathrm{HF}\rangle,
  \end{aligned}
  \label{eq:HW_dis}
\end{equation}
where the weights $w(\Omega_g)$ are determined by the quadrature methods chosen\cite{Xiasong2018_EfficientProjected,Shimizu2023_SO3Quadrature}.
The most efficient scheme is either a Lebedev quadrature for the angles $\alpha$ and $\beta$, and a periodic trapezoidal quadrature for the $\gamma$ angle (denoted as Lebedev-trapezoidal), or a periodic trapezoidal quadrature for $\alpha$ and $\gamma$, and a Gauss quadrature for $\cos\beta$  (denoted as Gauss-trapezoidal).
For a detailed recent study on quadrature rules for angular momentum projection see Ref.~\citenum{Shimizu2023_SO3Quadrature}.
The numerical integration can be easily parallelized over the quadrature points, since the matrix elements 
\begin{align}
    H(\Omega_g) &= \langle\Psi^\mathrm{HF}| \hat{H} \hat{R}(\Omega_g)| \Psi^\mathrm{HF}\rangle, 
    \label{eq:HamQuadPoint}\\
    W(\Omega_g) &= \langle\Psi^\mathrm{HF}| \hat{R}(\Omega_g)| \Psi^\mathrm{HF}\rangle,
    \label{eq:OvQuadPoint}
\end{align}
at each quadrature point, $g$, can be evaluated independently.
The final step for evaluating the matrix elements requires us to investigate how the rotation operator acts on the Hartree--Fock wave function and to evaluate the nonorthogonal matrix elements.
We will address these points in the next section.

\subsection{Rotation of basis functions}
\label{subsec:BasisRotation}

The Hartree--Fock wave function is composed of orbitals that are defined in terms of GTOs.
Hence, rotating the wave function requires evaluating the action of the rotation operator on a GTO.
For readability, we omit the particle type index $I$ in the following.
We rely on real-valued solid-harmonic GTOs, which are (up to the normalization factor) given as~\cite{Helgaker2014_Bible}
\begin{equation}
  G_{n\ell m}(\mathbf{r};a_n,\mathbf{s}_n) 
    = S_{\ell m}(\mathbf{r}-\mathbf{s}_n) \mathrm{e}^{-a_n(\mathbf{r}-\mathbf{s}_n)^2}\,,
  \label{eq:GTO}
\end{equation}
where $a_n$ is the Gaussian width, $\mathbf{s}_n$ is the shift, and $S_{\ell,m}$ is the real-valued solid harmonic, as defined, e.g., in Ref.~\citenum{Helgaker2014_Bible}. 
$n$ is the shell index while $\ell$ and $m$ are the angular momentum quantum numbers of the solid harmonics.
The rotation of the GTO reads
\begin{equation}
  \hat{R}(\Omega) {G}_{n\ell m}(\mathbf{r};a_n,\mathbf{s}_n) 
    = {G}_{n\ell m}(\mathbf{R}^{-1}(\Omega)\mathbf{r};a_n,\mathbf{s}_n)\,.
  \label{eq:GTO-Rotation}
\end{equation}
The rotation of the Gaussian function (i.e., the second term of the right-hand side of Eq.~(\ref{eq:GTO})) is straightforward to evaluate
\begin{equation}
  \mathrm{e}^{-a_n(\mathbf{R}^{-1}(\Omega)\mathbf{r}-\mathbf{s}_n)^2} 
    = \mathrm{e}^{-a_n(\mathbf{r}-\mathbf{R}(\Omega)\mathbf{s}_n)^2}\,,
  \label{eq:GTO-Rotation-2}
\end{equation}
since $\mathbf{R}^{-1}(\Omega)=\mathbf{R}^\mathrm{T}(\Omega)$.
Therefore, the rotated Gaussian is equivalent to the original Gaussian with a rotated center.
We rotate the solid harmonics as described in Ref.~\citenum{Pinchon2007_RotationSphericalHarmonics}.
We exploit the the matrix representation of a rotation in the solid harmonic basis $\mathbf{X}(\Omega)$ as defined in Ref.~\citenum{Pinchon2007_RotationSphericalHarmonics} to express the rotation of the GTO as
\begin{equation}
  \hat{R}(\Omega) {G}_{n\ell m}(\mathbf{r};a_n,\mathbf{s}_n)  = \sum_{\ell^\prime m^\prime} {G}_{n\ell^\prime m^\prime}(\mathbf{r};a_n,\mathbf{R}(\Omega)\mathbf{s}_{n}) X_{\ell^\prime m^\prime,\ell m}(\Omega).
  \label{eq:GTO-Rotation-Full}
\end{equation}
Next, we write a matrix element for an arbitrary one-body operator, $\hat{O}$, in the GTO basis with the center of the Gaussian in the ket rotated as
\begin{equation}
  \Tilde{O}_{n^\prime\ell^\prime m^\prime,n\ell m}(\Omega) = \langle {G}_{n^\prime\ell^\prime m^\prime}(\mathbf{s}_{n^\prime}) | \hat{O} | {G}_{n \ell m}(\mathbf{R}(\Omega)\mathbf{s}_n)\rangle \, ,
  \label{eq:RotatedMatrixElement}
\end{equation}
where we omitted the width of the Gaussian for readability. 
We now obtain the matrix element where both the center of the Gaussian and the solid harmonics are rotated according to the $\mathbf{X}$ matrix as follows
\begin{equation}
 \begin{aligned}
   O_{n^\prime\ell^\prime m^\prime,n^{\prime\prime}\ell^{\prime\prime}m^{\prime\prime}}(\Omega) &= \langle {G}_{n^\prime\ell^\prime m^\prime}(\mathbf{s}_{n^\prime}) | \hat{O} \hat{R}(\Omega) | {G}_{n^{\prime\prime}\ell^{\prime\prime}m^{\prime\prime}}(\mathbf{s}_{n^{\prime\prime}})\rangle \\
   &= \sum_{\ell m} \langle {G}_{n^\prime \ell^\prime m^\prime}(\mathbf{s}_{n^{\prime}}) | \hat{O} | {G}_{n^{\prime\prime} \ell m}(\mathbf{R}(\Omega)\mathbf{s}_{n^{\prime\prime}})\rangle X_{\ell m,\ell^{\prime\prime} m^{\prime\prime}}(\Omega).
 \end{aligned}
\end{equation}
We can rewrite this equation compactly in matrix notation where $\mathbf{X}(\Omega)$ is extended to the whole GTO space as $X_{n\ell m,n'\ell'm'} = \delta_{nn'} X_{\ell m,\ell'm'}$ such that
\begin{equation}
  O_{\mu\nu}(\Omega) = \sum_\kappa^L \Tilde{O}_{\mu\kappa}(\Omega) X_{\kappa\nu}(\Omega),
  \label{eq:FullMatrixOpertor}
\end{equation}
where the Greek indices are introduced as combined indices as $\mu=(n,\ell,m)$.
Consequently, for the rotation of a Hartree--Fock wave function, we first calculate the integrals in the GTO basis where the Gaussian shifts of the GTOs in the ket are rotated by $\Omega$.
Then, we evaluate the rotation matrix $\mathbf{X}(\Omega)$ and apply it to the integrals.
Two-body integrals can be rotated in complete analogy by rotating both Gaussian functions in the ket according to
\begin{equation}
  V_{\mu\nu\sigma\rho}(\Omega) = \sum_{\kappa\lambda}^L \Tilde{V}_{\mu\kappa\sigma\lambda}(\Omega) X_{\kappa\nu}(\Omega)X_{\lambda\rho}(\Omega),
  \label{eq:rotTwoBody}
\end{equation}
where $\Tilde{V}_{\mu\nu\sigma\rho}$ is a two-body integral in the chemistry notation over (possibly contracted) GTOs, $\chi_\mu(\mathbf{r})$, where the centers in the ket are rotated,
\begin{equation}
    \Tilde{V}_{\mu\nu\sigma\rho}(\Omega) = \int \int \mathrm{d}^3r_1\mathrm{d}^3r_2 
    G_\mu(\mathbf{r}_1; \mathbf{s}_\mu)
    G_\nu(\mathbf{r}_1; \mathbf{R}(\Omega)\mathbf{s}_\nu)
    g(\mathbf{r}_1,\mathbf{r}_2)
    G_\sigma(\mathbf{r}_2; \mathbf{s}_\sigma)
    G_\rho(\mathbf{r}_2; \mathbf{R}(\Omega)\mathbf{s}_\rho).
    \label{eq:rotatedTwoBodyIntegral}
\end{equation}
For the parity projection, $\hat{I}$ is applied to GTOs by inverting the origin of the GTO in such a way that $\mathbf{s}_n\rightarrow-\mathbf{s}_n$ and the solid-harmonics transform under the parity inversion as $(-1)^\ell$.
This operation can be easily incorporated into $\mathbf{X}(\Omega)$ by multiplying the diagonal with `$-1$' if $\ell$ is odd.

\subsection{Matrix elements and expectation values}
\label{subsec:MatrixElements}

We evaluate the matrix elements following the approach of Ref.~\citenum{Thom2009_Nonorthogonal}.
Note that the matrix elements cannot be evaluated with the strategy proposed in Refs.~\citenum{Jimenez2012_projected} and \citenum{Xiasong2018_EfficientProjected} for spin projection because a truncated GTO expansion does not constitute a complete basis in $\mathbb{R}^3$.
In fact, a GTO rotated by an arbitrary angle $\Omega$ cannot be represented, in general, in a finite linear combination of GTOs with a given fixed center.
We define the rotated GTO as
\begin{equation}
  \vert \phi_{Isj}(\Omega) \rangle = \hat{R}(\Omega) \vert \phi_{Isj} \rangle  ,
  \label{eq:RotatedOrbital}
\end{equation}
and the overlap matrix $B_{Isij}(\Omega)$ as
\begin{equation}
  B_{Isij}({\Omega}) = \langle\phi_{Isi}|\phi_{Isj}(\Omega)\rangle = \sum_{\mu\kappa\nu} C^*_{Is\mu i} \Tilde{S}_{I\mu\kappa}(\Omega)X_{\kappa\nu}(\Omega) C_{Is\nu j},
  \label{eq:MoOverlap}
\end{equation}
$\Tilde{S}_{I\mu\nu}(\Omega)$ is the overlap matrix in the GTO basis with the Gaussians in each ket rotated by $\Omega$.
The $ij$-cofactor of $\mathbf{B}_{Is}$ is defined as
\begin{equation}
  B_{Is}[i|j]({\Omega}) = (-1)^{i+j}\det \{ B_{Iskl}(\Omega) \}_{k\neq i,l\neq j}.
  \label{eq:Cofactor}
\end{equation}
We evaluate the matrix elements of the one-body contribution, $H_{1,I}$, for particles of type $I$ according to the L\"owdin rules \cite{Loewdin_NO_rules,Muolo2020_Nuclear} as
\begin{equation}
 \begin{aligned}  
  H_{1,I}(\Omega) &=  \langle\Psi^\text{HF}|\hat{H}_{1,I}\hat{R}(\Omega)|\Psi^\text{HF}\rangle  = \sum_{i}^{N_I} \langle\Psi^\text{HF}|\hat{h}_{I}(i)\hat{R}(\Omega)|\Psi^\text{HF}\rangle \\ 
  &= \prod_{J}^{N_\mathrm{t}}\prod_{s^\prime}^{S_J} \det \mathbf{B}_{Js^\prime}(\Omega)\sum_s^{S_I}\sum_{ij}^{N_{Is}} \langle\phi_{Isi}|\hat{h}_{I}|\phi_{Isj}(\Omega) \rangle (\mathbf{B}_{Is}^{-1}(\Omega))_{ji},
 \end{aligned}
 \label{eq:OneBodyMatElemIntermediate}
\end{equation}
where we have employed Cramer's rule
\begin{equation}
  (\mathbf{B}_{Is}^{-1}(\Omega))_{ji}   = \frac{B_{Is}[i|j]({\Omega})}{\det \mathbf{B}_{Is}(\Omega)} .
\end{equation}
Following the approach of Thom and Head-Gordon~\cite{Thom2009_Nonorthogonal}, we evaluate Eq.~(\ref{eq:OneBodyMatElemIntermediate}) in the the L\"owdin-paired basis,
\begin{equation}
  \langle^\mathrm{b}{\phi}_{Isi} | ^\mathrm{k}{\phi}_{Isj}(\Omega)\rangle = \sigma_{Isi}(\Omega) \delta_{ij} \, ,
  \label{eq:LowdinPair}
\end{equation}
where $\langle^\mathrm{b}{\phi}_{Isi}|$ are the bra orbitals and $|^\mathrm{k}{\phi}_{Isj}(\Omega)\rangle$ the ket orbitals.
This basis can be derived by a singular value decomposition of the overlap matrix
\begin{equation}
  \mathbf{B}_{Is}(\Omega) 
    = \mathbf{U}_{Is}(\Omega) \boldsymbol{\Sigma}_{Is}(\Omega) \mathbf{V}_{Is}^T(\Omega) \, ,
  \label{eq:OverlapPairing}
\end{equation}
where, $\boldsymbol{\Sigma}_{Is}(\Omega) = \mathrm{diag}(\sigma_{Is1}(\Omega),\dots)$ collects the singular values, while $\mathbf{U}_{Is}(\Omega)$ and $\mathbf{V}_{Is}(\Omega)$ denote the unitary transformation matrices defining the bra and rotated ket orbitals, respectively.
Note that the diagonalization of the inverse overlap matrix is obtained with the same transformation matrices as
\begin{equation}
    \mathbf{B}_{Is}^{-1}(\Omega) 
    = \mathbf{U}_{Is}(\Omega) \boldsymbol{\Sigma}^{-1}_{Is}(\Omega) \mathbf{V}_{Is}^T(\Omega) ,
  \label{eq:OverlapPairingInverse}
\end{equation}
which gives the inverse singular values.
$\mathbf{U}_{Is}$ defines a transformation that can be applied to the bra orbital coefficients as
\begin{equation}
  ^\mathrm{b} C_{Is\mu k} = \sum_{k}^{N_{Is}} U_{Isik}(\Omega) C_{Is\mu k},
  \label{eq:RotationBra}
\end{equation}
and, similarly, $\mathbf{V}_{Is}$ gives a transformation of the ket orbital coefficients as
\begin{equation}
  ^\mathrm{k} C_{Is\mu i} = \sum_{k}^{N_{Is}}\sum_{\nu}^{L_I}  V_{Isik}(\Omega) C_{Is\nu k} X_{I\nu\mu}(\Omega),
  \label{eq:RotationKet}
\end{equation}
where $\mathbf{X}_I(\Omega)=\{X_{I\mu\nu}(\Omega)\}$ is the solid harmonics rotation matrix introduced in Eq.~(\ref{eq:FullMatrixOpertor}).
Note that in order to be a phase-conserving orbital rotation, we exploit the fact that $\det(\mathbf{U}_{Is}) = \det(\mathbf{V}_{Is}(\Omega)) = 1$, which can be enforced by dividing the matrices by the value of their determinant.
Including the solid harmonics rotation matrix in the definition of the molecular orbital coefficients is advantageous as we do not need to first rotate the orbitals in the GTO basis and then calculate an expectation value, instead both can be achieved in a single step.
This is equivalent to including the orthonormalization matrix of the GTO overlap in the definition of the molecular orbitals which is the standard in the Hartree--Fock method.
Additionally, to write the matrix elements more compactly, we define
\begin{align}
  S_{Is}(\Omega) &= \prod_i^{N_Is} \sigma_{Isi}(\Omega) , \\
  S(\Omega)      &= \prod_I^{N_\mathrm{t}}\prod_s^{S_I} S_{Is}(\Omega) .
\end{align}
By inserting $\mathbbm{1}=\mathbf{U}^T_{Is}(\Omega)\mathbf{U}_{Is}(\Omega)$ and $\mathbbm{1}=\mathbf{V}^T_{Is}(\Omega)\mathbf{V}_{Is}(\Omega)$ in Eq.~(\ref{eq:OneBodyMatElemIntermediate}) allows us to simplify the matrix element as\cite{Thom2009_Nonorthogonal}
\begin{equation}
  H_{1,I}(\Omega) 
    = S(\Omega) \sum_s^{S_I}\sum_{i}^{N_{Is}} \langle^\mathrm{b}{\phi}_{Isi}|\hat{h}_{I}|^\mathrm{k}{\phi}_{Isi}(\Omega)\rangle 
      {\sigma^{-1}_{Isi}({\Omega})} \, ,
  \label{eq:OneBodyMatElem}
\end{equation}
where we employed
\begin{equation}
  (\boldsymbol{\Sigma}_{Is}^{-1})_{ji}(\Omega)  ={\sigma^{-1}_{Isi}({\Omega})}\delta_{ij}.
\end{equation}
To formulate the expectation value in the GTO basis, we introduce the codensity matrix as
\begin{equation}
  P_{Is\mu\nu}(\Omega) 
    = \sum_i^{N_{Is}}\frac{^\mathrm{b}{C}_{Is\mu i}(\Omega)^\mathrm{k}{C}_{Is\nu i} (\Omega)}{\sigma_{Isi}(\Omega)} \, ,
    \label{eq:codensity}
\end{equation}
which allows us to evaluate the matrix elements with the same procedure as for an orthonormal basis
\begin{equation}
  H_{1,I}(\Omega)
    = S(\Omega) \sum_s ^{S_I}\sum_{\mu\nu}^{L_I}  \Tilde{h}_{I\mu\nu}(\Omega) {P}_{Is\mu\nu}(\Omega) \, ,
\end{equation}
where $\Tilde{\mathbf{O}}_{I}(\Omega)$ is the GTO representation of the operator where the GTO centers are rotated in the ket.
The two-body matrix elements can be derived based on our previous works~\cite{Muolo2020_Nuclear,Feldmann2022_QuantumProton,Feldmann2023_Second} which we generalize here to a L\"owdin-paired basis. The two-body matrix elements for operators coupling particles of different types can again be evaluated with Cramer's rule. 
For the evaluation of matrix elements of operators coupling particles of the same type, we introduce the $ik,jl$-cofactor as
\begin{equation}
  B_{Is}[ik|jl](\Omega) = (-1)^{i+j+k+l} \det \{B_{Ismn}(\Omega)\}_{m\neq i,k; n\neq j,l},
  \label{eq:CofactorLowdin}
\end{equation}
which can be rewritten with the generalization of Cramer's rule\cite{Loewdin_NO_rules}
\begin{equation}
  \frac{B_{Is}[ik|jl](\Omega)}{S_{Is}(\Omega)} = \left(\mathbf{B}_{Is}^{-1}(\Omega)\right)_{ki}\left(\mathbf{B}_{Is}^{-1}(\Omega)\right)_{jl}-\left(\mathbf{B}_{Is}^{-1}(\Omega)\right)_{li}\left(\mathbf{B}_{Is}^{-1}(\Omega)\right)_{kj}.
\end{equation} 
To further evaluate all two-body matrix elements, we insert the identity in terms of the transformation matrices for both particles in analogy to Eq.~(\ref{eq:OneBodyMatElem}) and insert the definition of the codensity matrix of Eq.~(\ref{eq:codensity}).
Consequently, we can write the Hamiltonian matrix element, Eq.~(\ref{eq:HamQuadPoint}), at a quadrature point as
\begin{equation}
 \begin{aligned}
    H(\Omega)  = S(\Omega)\Bigg(\ 
        & \sum^{N_\mathrm{t}}_I \sum_s^{S_I} \sum_{\mu\nu}^{L_I} \Tilde{h}_{I\mu\nu}(\Omega) P_{Is\mu\nu}(\Omega) \\
        & + \frac{1}{2} \sum^{N_\mathrm{t}}_I \sum_s^{S_I} \sum_{\mu\nu\rho\sigma}^{L_I} 
            P_{Is\mu\nu} (\Omega)\Tilde{V}^\mathrm{a}_{I\mu\nu,I\rho\sigma}(\Omega) P_{Is\sigma\rho}(\Omega) \\
        & + \sum^{N_\mathrm{t}}_I \sum_{s<t}^{S_I} \sum_{\mu\nu\rho\sigma}^{L_I}
            P_{Is\mu\nu}(\Omega) \Tilde{V}_{I\mu\nu,I\rho\sigma}(\Omega) P_{It\sigma\rho}(\Omega) \\
        & + \sum^{N_\mathrm{t}}_{I<J} \sum_{s}^{S_I} \sum_{t}^{S_J}\sum_{\mu\nu}^{L_I} \sum_{\rho\sigma}^{L_J}
            P_{Is\mu\nu}(\Omega) \Tilde{V}_{I\mu\nu,J\rho\sigma}(\Omega) P_{Jt\sigma\rho}(\Omega)\Bigg).
 \end{aligned}
 \label{eq:NuclearElectronicEnergy}
\end{equation}
Here, $\Tilde{h}_{I\mu\nu}(\Omega)$ denotes the one-body matrix elements in the GTO basis with the center of the GTO rotated in the ket, as in Eq.~(\ref{eq:RotatedMatrixElement}), and $\Tilde{V}_{I\mu\nu,J\rho\sigma}(\Omega)$ refers to the two-body matrix element where the centers of both GTOs in the ket are rotated, as in Eq.~(\ref{eq:rotatedTwoBodyIntegral}).
The superscript in $\Tilde{V}^\mathrm{a}_{I\mu\nu,I\rho\sigma}(\Omega)$ indicates that the integral is anti-symmetrized.
The overlap matrix element, Eq.~(\ref{eq:OvQuadPoint}), is given by $W(\Omega) = S(\Omega)$.
Note that the computational scaling of evaluating the expectation value of the Hamiltonian in the L\"owdin-paired basis can be reduced from $\mathcal{O}(L^4N^4)$ to $\mathcal{O}(L^4)$.
The former scaling stems from the fact that the two-body integrals have to be transformed to the molecular orbital basis with the integral direct method.
Conversely, in the L\"owdin-paired basis, the contraction of the two-body integrals with the codensity can be carried out equivalently to the unprojected Hartree--Fock theory.\cite{Almlof1982_SAD} 
Consequently, the computational scaling at each quadrature point is the same as in unprojected NE-HF.\cite{Feldmann2022_QuantumProton,Feldmann2023_Second}
At each quadrature point, we evaluate Eq.~(\ref{eq:NuclearElectronicEnergy}) with an integral-direct method.
Finally, we solve the Hill--Wheeler equations that read, in matrix form, 
\begin{equation}
  \mathbf{H}^J \mathbf{f} = E^J \mathbf{W}^J\mathbf{f},\quad E^J\in\mathbb{R},\ \mathbf{f}\in\mathbb{C}^{2J+1},\ \mathbf{H}^J, \mathbf{W}^J\in \mathbb{C}^{2J+1\times 2J+1},
  \label{eq:MatrixFormWheeler}
\end{equation}
where $\mathbf{H}^J$ and $\mathbf{W}^J$ are hermitian and $\mathbf{W}^J$ can have eigenvalues $\lambda_k\geq0$.
As a consequence, $\mathbf{H}^J$ has only as many eigenvalues as $\mathbf{W}^J$ has non-zero eigenvalues.
We solve the generalized eigenvalue problem in Eq.~(\ref{eq:MatrixFormWheeler}) with the canonical L\"owdin orthonormalization method (see, e.g., Ref.~\citenum{Muolo2020_Nuclear}) that is also commonly used to solve the Roothaan--Hall equations: we first diagonalize $\mathbf{W}^J$ and discard all eigenvalues and their eigenvectors that are below a given numerical threshold.
Then, assuming there are $k$ non-zero eigenvalues, we set up the transformation matrix $\mathbf{Y} \in \mathbb{R}^{2J+1,k}$ that orthonormalizes the overlap as $\mathbf{Y}^T\mathbf{W}^J\mathbf{Y}=\mathbbm{1}_{k\times k}$, such that we can transform the Hill--Wheeler equations into a hermitian standard eigenvalue problem
\begin{equation}
  \Tilde{\mathbf{H}}^J \Tilde{\mathbf{f}} = E^J \Tilde{\mathbf{f}} \, ,
  \label{eq:HW_orthonormal}
\end{equation}
where $\Tilde{\mathbf{H}}^J = \mathbf{Y}^T\mathbf{H}^J\mathbf{Y}$.

\subsection{Spin expectation value}
\label{sec:app_s2}

We conclude the implementation section by deriving the expectation value of the $\hat{S}^2$ operator, 
\begin{equation}
  \hat{S}^2 = \sum_i^N \hat{s}_{z,i} + \sum_i^N \hat{s}_{z,i}^2 + \sum_{i\neq j}^N \hat{s}_{z,i}\hat{s}_{z,j} + \sum_i^N \hat{s}_{-,i}\hat{s}_{+,i} + \sum_{i\neq j}^N \hat{s}_{-,i}\hat{s}_{+,j},
  \label{eq:S2}
\end{equation}
which can be calculated efficiently within the L\"owdin-paired basis.
For clarity, we focus here on a single-particle-type wave function with spin-orbitals and evaluate the matrix elements between $|\Phi\rangle$ with basis functions $\{|\phi_i\rangle\}$ and $|\Psi\rangle$ with basis functions $\{|\psi_i\rangle\}$, and the overlap is $B_{ij} = \braket{\phi_i | \psi_j}$.
We first introduce the spin-specific overlap
\begin{equation}
  S_s = \prod_{i}^{N_s} \sigma_{s i},
  \label{eq:SpinSpecificOverlap}
\end{equation}
and the reduced overlap vector $a^s_i$
\begin{equation}
  a^s_i = \prod_{j\neq i}^{N_s} \sigma_{s j}.
  \label{eq:ReducedOverlap}
\end{equation}
With those, we can express the expectation value of the first term in Eq.~(\ref{eq:S2}) as
\begin{equation}
  \sum_i^N \langle\Phi | \hat{s}_{z,i} | \Psi \rangle = \frac{1}{2} \Big( S_\downarrow \sum_{i}^{N_\uparrow} \langle \phi_{\uparrow i} | \psi_{\uparrow i} \rangle a^\uparrow_i
    - S_\uparrow \sum_{i}^{N_\downarrow} \langle \phi_{\downarrow i} | \psi_{\downarrow i} \rangle a^\downarrow_i. \Big)
  \label{eq:SpinExpectationValue}
\end{equation}
For the second term, we have
\begin{equation}
  \sum_i^N \langle\Phi | \hat{s}_{z,i}^2 | \Psi \rangle = \frac{1}{4} \Big( S_\downarrow \sum_{i}^{N_\uparrow} \langle \phi_{\uparrow i} | \psi_{\uparrow i} \rangle a^\uparrow_i
   + S_\uparrow \sum_{i}^{N_\downarrow} \langle \phi_{\downarrow i} | \psi_{\downarrow i} \rangle a^\downarrow_i \Big).
\end{equation}
In contrast to the other terms, the expectation value of the third term is evaluated in the spin-orbital basis and is given by
\begin{equation}
\sum_{i\neq j}^N \langle\Phi | \hat{s}_{z,i}\hat{s}_{z,j} | \Psi \rangle = \sum_{i\neq j}^{N} \langle \phi_{i} | \hat{s}_{z,i} | \psi_{i} \rangle \langle \phi_{j} | \hat{s}_{z,j} | \psi_{j} \rangle \prod_{k\neq i,j}\sigma_{k}.
\end{equation}
The fourth term can be evaluated according to
\begin{equation}
    \sum_i^N \langle\Phi | \hat{s}_{-,i}\hat{s}_{+,i}  | \Psi \rangle =  S_\uparrow \sum_{i}^{N_\downarrow} \langle \phi_{\downarrow i} | \psi_{\downarrow i} \rangle a_i^\downarrow,
\end{equation}
and, finally, the expectation value of the final term is expressed as
\begin{equation}
    \sum_{i\neq j}^N \langle\Phi | \hat{s}_{-,i}\hat{s}_{+,j}| \Psi \rangle = - \sum_{i}^{N_\downarrow} \sum_j^{N_\uparrow} 
    \langle \phi_{\downarrow i} | \psi_{\uparrow j} \rangle  \langle \phi_{\uparrow j} | \psi_{\downarrow i} \rangle a^\uparrow_j a^\downarrow_i,
\end{equation}
which completes the evaluation of the expectation value of the $\hat{S}^2$ operator within the L\"owdin-paired basis.

\section{Computational details}
\label{sec:ComputationalDetails}

For this work, we employed the correlation-consistent (cc) basis sets~\cite{Dunning1989_gaussian} for electrons, augmented with specialized functions for multicomponent (mc) calculations developed by Brorsen et al.\ \cite{Brorsen2023_BasisSets}.
For the protons, we chose the protonic basis (PB) by Hammes-Schiffer and coworkers \cite{Hammes-Schiffer2020_NuclearBasis}.
We implemented the projected NE-HF method in the open-source Kiwi program \cite{kiwi}.
We calculated the integrals with our integral evaluation package \cite{integralevaluator} which relies on Libint \cite{Libint2}. 
For H$_2$, we set the bond length to 0.74~{\AA} while, for H$_3^+$, we adopt an equilateral triangular structure with an edge length of 0.867850~{\AA}. 
Moreover, we calculated the classical center of mass of the system based on the centers of the Gaussians and we transform the coordinates such that the center of mass is in the origin of the coordinate system.
For H$_2$, the center of mass is the center of the bond, and for H$_3^+$, it is the center of the equilateral triangle.
Note that the overlap matrix om Eq.~(\ref{eq:MoOverlap}) may be singular.
However, a singular overlap matrix indicates that the corresponding quadrature point yields a null-energy contribution.
We, therefore, neglected quadrature points for which the overlap matrix contains singular values with an absolute value smaller than a numerical threshold of $10^{-5}$ in atomic units.
The matrix elements $H(\Omega_g)$ and $W(\Omega_g)$ can be evaluated independently for different quadrature points, $g$.
We, therefore, parallelized the construction of the $\mathbf{H}^J$ and $\mathbf{W}^J$ matrices via shared-memory OpenMP parallelization.
Specifically, each thread evaluates $H(\Omega_g)$ and $W(\Omega_g)$ for a subset of quadrature points and accumulates the result into thread-specific local $\mathbf{H}^J$ and $\mathbf{W}^J$ matrices.
The local matrices of each thread are then added to construct the full Hill--Wheeler matrices.

\section{Results and Discussion}
\label{sec:results}

\subsection{Convergence of the quadrature}

We first investigated the convergence of the symmetry-projected NE-HF energy with the number of quadrature points of the numerical integration for the H$_2$ molecule in the $J=0$ and $M_{S,\pr}=0$ state.
As we mentioned above, the action of the projection operator on the Hartree--Fock wave function depends on the orientation of the symmetry-broken Hartree--Fock wave function.
Therefore, the numerical integration can be more or less efficient depending on the positions of the centers of the GTOs.
Not surprisingly for a linear molecule, we found that orienting the basis functions along the $z$-axis leads to the fastest energy convergence.
Moreover, in this case, the combination of trapezoidal and Gauss quadrature converged faster than the Lebedev trapezoidal quadrature. 
The convergence of the ground state energy with the number of quadrature points per angle is shown in Table~\ref{tab:h2_gtid}.
The ground state energy was converged at least up to $1\ \mathrm{\upmu Ha}$ with $N_q=20$ points per angle, yielding $N_q^3=8000$ quadrature points in total.

\begin{table}
  \centering 
  \caption{Convergence of the energy in atomic units with the number of quadrature points per angle, $N_q$, and the total number of quadrature points $N^3_q$, of the Gauss-trapezoidal quadrature for the H$_2$ system.
  We fix the proton basis set to PB4-D and vary the electronic basis set.} 
  \label{tab:h2_gtid}
  \begin{tabular}{c c c c c }
    \hline\hline
    $N_q$ & $N_q^3$  &      cc-pVDZ-mc     &    cc-pVTZ-mc    &      cc-pVQZ-mc   \\ 
    \hline                         
    10  & 1000      & $-1.096\ 932$ & $-1.097\ 670$ & $-1.097\ 793$ \\
    15  & 3375      & $-1.095\ 345$ & $-1.096\ 069$ & $-1.096\ 171$ \\
    20  & 8000      & $-1.095\ 341$ & $-1.096\ 061$ & $-1.096\ 165$ \\ 
    25  & 15625      & $-1.095\ 341$ & $-1.096\ 061$ & $-1.096\ 165$ \\
    30  & 27000      & $-1.095\ 341$ & $-1.096\ 061$ & $-1.096\ 165$ \\
    \hline\hline
  \end{tabular} 
\end{table}

\subsection{Absolute and relative energies of H$_2$}
\label{sec:H2}

We investigated the convergence of the absolute energies and energy differences between the $J=0,M_{S,\pr}=0$ and $J=1,M_{S,\pr}=1$ states given in Table~\ref{tab:h2_basisSet}.
The absolute energy converges with increasing electronic basis set size, but no convergence is observed with the protonic basis set size.
We already discussed this effect in our previous work~\cite{Feldmann2022_QuantumProton} where we ascribed the lack of convergence to the fact that the PB basis sets were optimized only for correlated calculations. 
Table~\ref{tab:h2_basisSet} also contains that the energy differences with different combinations of basis sets are in the range of $2\ \mathrm{cm}^{-1}$, and hence, we may conclude that the energy difference is converged already with the smallest basis set combination.

\begin{table}
  \centering 
  \caption{Convergence of the energy with the electronic (cc-pVDZ-mc, cc-pVTZ-mc, and cc-pVQZ-mc) and protonic (PB4-D, PB4-F1, and PB5-G) basis set for the $J=0$ and $J=1$ states and for the transition energy $\Delta E_{1 \leftarrow 0}$ of H$_2$.
  Absolute energies are given in atomic units and relative energies in cm$^{-1}$.
  We chose the Gauss-trapezoidal quadrature with 20 points for each angle, resulting in 8000 quadrature points in total.\\
  * 25 quadrature points for each angle and 15625 points overall were required to reach convergence for the largest basis set combination of cc-pVQZ-mc and PB5-G.} 
  \label{tab:h2_basisSet}
  \begin{tabular}{l c c c } 
  \hline\hline
  p$^+$/e$^-$    &      cc-pVDZ-mc     &    cc-pVTZ-mc    &      cc-pVQZ-mc   \\ 
  \hline     
  $E_{\mathrm{HF}}$  &               &               &                             \\ 
  \hline                         
  PB4-D          & $-1.073\ 789$ & $-1.074\ 474$ & $-1.074\ 572$ \\
  PB4-F1         & $-1.073\ 752$ & $-1.074\ 437$ & $-1.074\ 534$ \\
  PB5-G          & $-1.073\ 842$ & $-1.074\ 525$ & $-1.074\ 626$ \\
  \hline     
  $E_{J=0,S_{z,\pr}=0}$  &               &               &                             \\ 
  \hline                         
  PB4-D          & $-1.095\ 341$ & $-1.096\ 061$ & $-1.096\ 139$ \\
  PB4-F1         & $-1.095\ 273$ & $-1.095\ 990$ & $-1.096\ 068$ \\
  PB5-G          & $-1.095\ 436$ & $-1.096\ 153$ & $-1.096\ 123^*$ \\
  \hline     
  $E_{J=1,S_{z,\pr}=1}$   &               &               &                                \\ 
  \hline                          
  PB4-D          & $-1.094\ 685$ & $-1.095\ 406$ & $-1.095\ 485$ \\
  PB4-F1         & $-1.094\ 617$ & $-1.095\ 337$ & $-1.095\ 412$ \\
  PB5-G          & $-1.094\ 789$ & $-1.095\ 495$ & $-1.095\ 577^*$ \\
  \hline                          
  $\Delta E_{1 \leftarrow 0}/$cm$^{-1}$ &         &               &                            \\
  \hline                          
  PB4-D          & 144.12  & 143.91 & 143.66 \\ 
  PB4-F1         & 144.00  & 143.48 & 144.04 \\ 
  PB5-G          & 142.18  & 144.42 & 142.96$^*$ \\
  \hline\hline
\end{tabular} 
\end{table}
Next, we compare the energy difference between rotational states to the value calculated by Pachucki and Komasa$^{55}$, which was obtained based on the Born--Oppenheimer approximation where the potential energy surface was generated with a wave function composed of explicitly-correlated Gaussians and the one-dimensional nuclear equation was solved numerically on a grid. Compared to their result of $118.55\ \mathrm{cm}^{-1}$, we find a deviation of $24.41\ \mathrm{cm}^{-1}$ to our energy difference of $142.96\ \mathrm{cm}^{-1}$. We note that, since the nuclear Schr\"odinger equation arising from the Born--Oppenheimer approximation was solved numerically on a grid in the reference work, no projection onto rotational states was necessary.

Given the well-known limitations of the NE-HF method, as previously discussed in the literature~\cite{Hammes-Schiffer2020_Review,Feldmann2022_QuantumProton}, the observed deviation is likely due to the missing correlation energy.
Another contributing factor could be that our method optimizes first the NE-HF wave function, and then projects it onto the rotational state of interest (known as projection-after-variation) without accounting for orbital relaxation.
This effect is included in a variation-after-projection approach that is computationally more demanding and it requires the derivation of gradients of the projected energy with respect to the orbital coefficients or orbital rotations \cite{ring_2004nuclear}.

An alternative approach for removing rotational contributions from NE-HF has been proposed by Nakai et al.~\cite{Nakai2005_EliminationRotTrans}
Within their approach, an approximate correction term, based on expanding the rigid-rotor rotational energy around an equilibrium geometry, is included to remove rotational energy contribution from the Hamiltonian.
The equilibrium geometry, however, is introduced a posteriori and it breaks the nuclear indistinguishability.
The resulting Hartree--Fock energy without rotational energy is lower by $29.774\ \mathrm{mHa}$ compared to the energy including the rotational contribution.
With our method, we observe a lowering of the energy by $21.497\ \mathrm{mHa}$. The deviation to Nakai's result is $8.277\ \mathrm{mHa}$, which was obtained with a different basis set.

We emphasize that the value of $21.497\ \mathrm{mHa}$ is equivalent to $4718.05\ \mathrm{cm}^{-1}$.
Considering the rotational constant of $59.275\ \mathrm{cm}^{-1}$, this finding indicates that very high rotational states contribute to the symmetry-broken wave function.
Moreover, Nakai's approach includes orbital relaxation effects because it removes rotational contributions from the Hamiltonian that is used in the variational minimization.
This suggests that the remaining energy lowering may be due to the missing orbital relaxation effects which could be included with a variation-after-projection scheme.
However, while the approach of Ref.~\citenum{Nakai2005_EliminationRotTrans} can target only the rotational ground state, our algorithm can target rotational states with arbitrary $J$ values.
The possibility of targeting rotationally excited states is a key advantage that makes our method applicable to high-resolution spectroscopy calculations.
This will allow for the computation of rotational spectra which is not possible when the rotational motion of the system is eliminated from the Hamiltonian.

A second advantage is the possibility of fully controlling the rotational symmetry properties of the wave function, which allows us to capture essential physics.
For instance, in the H$_2$ molecule, the $J=0$ state is Pauli-forbidden for $S^2_{\pr}=3$.
If we attempt to project onto this state in our calculations, we will observe that it is forbidden since the $\mathbf{W}^J$ matrix is numerically zero, i.e., the projector projects onto the null vector.
The same holds true for the $J=1$ state, which is forbidden for $S^2_{\pr}=0$.
Additionally, the NE-HF wave function for $M_{S,\pr}=0$ exhibits broken symmetry in the nuclear spin.
By implementing the spin expectation value for the projected NE-HF wave function, we observed that in this case, spin symmetry is exactly restored as a byproduct of the symmetry projection.
The spin expectation values were evaluated according to the equations provided in Appendix~\ref{sec:app_s2}.

Lastly, we have calculated the overlap of the symmetry-broken Hartree--Fock wave function with the projected Hartree--Fock wave function for the ground state with the cc-pVTZ-mc and PB4-D basis sets. The numerical evaluation yielded a value of $\langle\Psi_\mathrm{HF} | \Psi^{\mathrm{PHF}}_{0,0,1} \rangle = 0.087\ 839$. Hence, the symmetry-broken wave function barely overlaps with the projected wave function. The fact that a Hartree--Fock wave function is considered qualitatively inaccurate in electronic structure theory when the overlap with the full-CI wave function (which corresponds to the $C_0$ coefficient) is less than, say, 0.9 signifies that the symmetry broken nuclear-electronic wave function does not resemble the correct wave function at all.

\subsection{Rotational states of H$_3^+$}

In this section, we investigate absolute and relative energies of the H$_3^+$ molecule where also parity must be considered. 
Based on our findings from Sec.~\ref{sec:H2}, which show no significant effect on relative energies for the chosen basis set combinations, we selected the combination of the PB4-D and cc-pVTZ-mc basis sets for all calculations on H$_3^+$.
We centered the Gaussian functions of one hydrogen atom on the $z$-axis. We analyzed the energy convergence with the quadrature method and found that the combination of the Lebedev and trapezoidal rule yields faster convergence than the Gauss-trapezoidal quadrature.
The quadrature is fully converged with 30 quadrature points for the trapezoidal rule and with 434 quadrature points for the Lebedev rule, resulting in 13020 points in total. 

The $J=0$ state of the H$_3^+$ molecule is well-known to be Pauli-forbidden and, therefore, the ground state is the $J=1$ state~\cite{Lindsay2001_ComprehensiveH3plus,muolo2019}.
Our calculations also corroborate this selection rule, as projecting onto the $J=0$ state for all parity and nuclear spin combinations yields an overlap, $\mathbf{W}^0$, that is in all cases smaller than $10^{-6}$ atomic units.

We calculated the energy of the first seven states of the molecule, taking into account their proton spin, angular momentum, and parity quantum numbers.
As we can only fix $M_{S,\pr}$ in the Hartree--Fock wave function, we calculated the $S_{\pr}^2$ expectation value.
We sorted the states according to the energy, calculated the energy differences, and compared them to experimental reference values from Ref.~\citenum{Lindsay2001_ComprehensiveH3plus}.
The results presented in Table~\ref{tab:h3p_energy_levels} show that our calculations successfully reproduce the energy ordering when compared to the reference, while also confirming that $S_{\pr}^2$ is exactly restored.
Moreover, although the absolute deviation of the energy differences from the reference increases with rising energy, the relative error remains approximately constant.

\begin{table}
  \centering 
  \caption{Low-lying rotational states of the H$_3^+$ molecule.
  The $S_{z,\pr}$ quantum number is exactly specified in the Hartree--Fock wave function, the quantum numbers $J$ and $p$ are restored by our projection procedure, while $\braket{S^2_{\pr}}$ is not guaranteed to be conserved and is, therefore, evaluated as an expectation value.
  We rely on the cc-pVTZ-mc basis set for the electrons and the PB4-D basis set for the protons and we applied the Lebedev-trapezoidal rule with 13020 quadrature points.
  The reference energies, $\Delta E_\mathrm{ref}$, are taken from Ref.~\citenum{Lindsay2001_ComprehensiveH3plus}.} 
  \label{tab:h3p_energy_levels}
  \begin{tabular}{c c c c c c c} 
  \hline\hline
  $J$ & $M_{S,{\pr}}$ & $p$ & $\braket{S^2_{\pr}}$ &  $E/\mathrm{Ha}$ & $\Delta E /\mathrm{cm}^{-1}$ & $\Delta E_\mathrm{ref} /\mathrm{cm}^{-1}$\\\hline
  1 & ${1}/{2}$ & $-1$ & $ 0.7500$ & $-1.236\ 663$  & 0        & 0 \\
  1 & ${3}/{2}$ & $1$ & $ 3.7500$ & $-1.236\ 524$  & 30.59    & 22.84 \\
  2 & ${1}/{2}$ & $1$  & $ 0.7500$ & $-1.236\ 072$  & 129.71   & 105.17 \\
  2 & ${1}/{2}$ & $-1$ & $ 0.7500$ & $-1.235\ 652$  & 221.89   & 173.23 \\
  3 & ${3}/{2}$ & $-1$ & $ 3.7500$ & $-1.235\ 256$  & 308.77   & 251.22 \\
  3 & ${1}/{2}$ & $-1$ & $ 3.7499$ & $-1.235\ 255$  & 309.24   & 251.22 \\
  3 & ${1}/{2}$ & $1$  & $ 0.7500$ & $-1.234\ 554$  & 462.87   & 363.89 \\
  3 & ${1}/{2}$ & $-1$ & $ 0.75$ & $-$  & $-$   & $437.90$ \\
  \hline\hline
\end{tabular} 
\end{table}

The results in Table~\ref{tab:h3p_energy_levels} indicate that we have obtained two states with identical energy.
Specifically, when targeting the $J=3, p=-1$ state, we set $M_{S,\pr}=\frac{1}{2}$ and $M_{S,\pr}=\frac{3}{2}$, which both resulted in the same energy and $S_{\pr}^2$ expectation value.
Consequently, for both $M_{S,\pr}$ values, the projection leads to the same state.
In those cases where only one $S_{\pr}^2$ state exists within the subspace defined by specific $J$ and $p$ quantum numbers, the projection automatically targets the desired $S_{\pr}^2$ state.
However, if multiple $S_{\pr}^2$ values are possible for a $(J,p)$ pair, we cannot target a specific $S_{\pr}^2$ state because only the $M_{S,\pr}$ quantum number can be specified in the Hartree--Fock ansatz.
To target the highest energy $J=3$ state where $S_{\pr}^2=0.75$, which is the last line of Table~\ref{tab:h3p_energy_levels}, we would require an additional projection onto eigenstates of the spin operator.

Finally, since our approach is not a cheap method, we report the timings of the H$_3^+$ projection to address the computational demands of the method: the numerical projection carried out on 64 cores on an Intel(R) Xeon(R) Gold 6136 central processing unit required 1 hour and 15 minutes.
  Hence, in practice, the timings are modest on modern computing nodes because of the very efficient parallelization. 
  Moreover, for the first approach with non-exponential scaling to target specific rotational states without requiring any prior knowledge of the system, this appears acceptable.
  We also note that
  the number of single-point calculations required for ro-vibrtaional structure calculations based on potential energy surfaces is similar, if not higher, compared to the number of quadrature points in our approach.

\section{Conclusions}

We presented a method for projecting the nuclear-electronic Hartree--Fock wave function onto states with specific rotational and parity quantum numbers.
This allows for removing contamination arising from rotationally excited states and, consequently, for obtaining a better variational estimate of the ground-state energy.
In contrast to the approximate scheme introduced in Ref.~\citenum{Nakai2005_EliminationRotTrans}, our method eliminates rotational contributions exactly, thereby addressing the limitations that were highlighted by Sutcliffe~\cite{Sutcliffe2005_CommentNakai}.
Furthermore, our ansatz fulfills the requirements recently proposed by Sutcliffe~\cite{Sutcliffe2021_Respect}, allowing for the exact treatment of permutational and rotational symmetry of nuclei, which is lacking in other orbital-based pre-Born--Oppenheimer methods.
We demonstrated the reliability of our approach for the prototypical molecules H$_2$ and H$_3^+$ by comparing our results to experimental data and the results obtained by Nakai et al.~\cite{nakai2007nuclear}.

This work lays the foundation for further improvements of the NE-HF and related methods. 
First, a variation-after-projection approach that variationally optimizes the Hartree--Fock wave function in the presence of the projection operator would allow for including orbital relaxation effects and, therefore, further improve the accuracy of the NE-HF energy \cite{Jimenez2012_projected}.
Although correlation effects can be considered by post-Hartree--Fock approaches\cite{Brorsen2020_SelectedCI-PreBO,Pavosevic2019_MulticomponentCC,Muolo2020_Nuclear,Feldmann2022_QuantumProton,Pavosevic_OOMP2-Multicomponent}, this will significantly increase the computational costs.
A good balance between costs and accuracy can be achieved within nuclear-electronic density functional theory as demonstrated by Hammes-Schiffer and coworkers\cite{Hammes-Schiffer2012_MulticomponentDFT,Hammes-Schiffer2017_MulticomponentDFT,Brorsen2018_TransferableDFT,HammesSchiffer2019_GGA-DFT} to which our approach developed here can be directly applied.
Finally, our method can be extended to molecules containing heavy atoms beyond H since the computational scaling of $\mathcal{O}(N^4)$ in the number of orbitals is not prohibitive. Moreover, parallelization is highly efficient. However, to this end, atom-centered nuclear basis functions must be optimized for heavier nuclei based, e.g., on the algorithm described in Ref.~\citenum{Hammes-Schiffer2020_NuclearBasis}.

\section*{Acknowledgments}

R.~F. gratefully acknowledges financial support by the G\"unthard Foundation for a PhD scholarship.
M.~R. and A.~B. are grateful for generous funding through the ``Quantum for Life Center'' funded by the Novo Nordisk Foundation (grant NNF20OC0059939).



\begin{mcitethebibliography}{59}
\providecommand*\natexlab[1]{#1}
\providecommand*\mciteSetBstSublistMode[1]{}
\providecommand*\mciteSetBstMaxWidthForm[2]{}
\providecommand*\mciteBstWouldAddEndPuncttrue
  {\def\EndOfBibitem{\unskip.}}
\providecommand*\mciteBstWouldAddEndPunctfalse
  {\let\EndOfBibitem\relax}
\providecommand*\mciteSetBstMidEndSepPunct[3]{}
\providecommand*\mciteSetBstSublistLabelBeginEnd[3]{}
\providecommand*\EndOfBibitem{}
\mciteSetBstSublistMode{f}
\mciteSetBstMaxWidthForm{subitem}{(\alph{mcitesubitemcount})}
\mciteSetBstSublistLabelBeginEnd
  {\mcitemaxwidthsubitemform\space}
  {\relax}
  {\relax}

\bibitem[Pavo{\v{s}}evi{\'c} \latin{et~al.}(2020)Pavo{\v{s}}evi{\'c}, Culpitt,
  and Hammes-Schiffer]{Hammes-Schiffer2020_Review}
Pavo{\v{s}}evi{\'c},~F.; Culpitt,~T.; Hammes-Schiffer,~S. {Multicomponent
  Quantum Chemistry: Integrating Electronic and Nuclear Quantum Effects via the
  Nu\-clear--Electronic Orbital Method}. \emph{Chem. Rev.} \textbf{2020},
  \emph{120}, 4222--4253\relax
\mciteBstWouldAddEndPuncttrue
\mciteSetBstMidEndSepPunct{\mcitedefaultmidpunct}
{\mcitedefaultendpunct}{\mcitedefaultseppunct}\relax
\EndOfBibitem
\bibitem[Born and Oppenheimer(1927)Born, and Oppenheimer]{Born-Oppenheimer1927}
Born,~M.; Oppenheimer,~R. Zur Quantentheorie der Molekeln. \emph{Ann. Phys.}
  \textbf{1927}, \emph{389}, 457--484\relax
\mciteBstWouldAddEndPuncttrue
\mciteSetBstMidEndSepPunct{\mcitedefaultmidpunct}
{\mcitedefaultendpunct}{\mcitedefaultseppunct}\relax
\EndOfBibitem
\bibitem[Braams and Bowman(2009)Braams, and Bowman]{Bowman2009_Permutationally}
Braams,~B.~J.; Bowman,~J.~M. Permutationally invariant potential energy
  surfaces in high dimensionality. \emph{Int. Rev. Phys. Chem.} \textbf{2009},
  \emph{28}, 577--606\relax
\mciteBstWouldAddEndPuncttrue
\mciteSetBstMidEndSepPunct{\mcitedefaultmidpunct}
{\mcitedefaultendpunct}{\mcitedefaultseppunct}\relax
\EndOfBibitem
\bibitem[Manzhos and Carrington~Jr.(2020)Manzhos, and
  Carrington~Jr.]{Carrington2020_NeuralNetworkPES}
Manzhos,~S.; Carrington~Jr.,~T. Neural network potential energy surfaces for
  small molecules and reactions. \emph{Chem. Rev.} \textbf{2020}, \emph{121},
  10187--10217\relax
\mciteBstWouldAddEndPuncttrue
\mciteSetBstMidEndSepPunct{\mcitedefaultmidpunct}
{\mcitedefaultendpunct}{\mcitedefaultseppunct}\relax
\EndOfBibitem
\bibitem[Thomas(1969)]{Thomas1969_protonic1}
Thomas,~I.~L. Protonic structure of molecules. I. Ammonia molecules.
  \emph{Phys. Rev.} \textbf{1969}, \emph{185}, 90\relax
\mciteBstWouldAddEndPuncttrue
\mciteSetBstMidEndSepPunct{\mcitedefaultmidpunct}
{\mcitedefaultendpunct}{\mcitedefaultseppunct}\relax
\EndOfBibitem
\bibitem[Thomas(1969)]{Thomas1969_protonic2}
Thomas,~I.~L. {The protonic structure of methane, ammonia, water, and hydrogen
  fluoride}. \emph{Chem. Phys. Lett.} \textbf{1969}, \emph{3}, 705--706\relax
\mciteBstWouldAddEndPuncttrue
\mciteSetBstMidEndSepPunct{\mcitedefaultmidpunct}
{\mcitedefaultendpunct}{\mcitedefaultseppunct}\relax
\EndOfBibitem
\bibitem[Thomas and Joy(1970)Thomas, and Joy]{Thomas1970_protonic3}
Thomas,~I.~L.; Joy,~H.~W. {Protonic Structure of Molecules. II. Methodology,
  Center-of-Mass Transformation, and the Structure of Methane, Ammonia, and
  Water}. \emph{Phys. Rev. A} \textbf{1970}, \emph{2}, 1200\relax
\mciteBstWouldAddEndPuncttrue
\mciteSetBstMidEndSepPunct{\mcitedefaultmidpunct}
{\mcitedefaultendpunct}{\mcitedefaultseppunct}\relax
\EndOfBibitem
\bibitem[Thomas(1971)]{Thomas1971_protonic4}
Thomas,~I.~L. ``Vibrational'' and ``Rotational'' Energy Levels as Protonic
  Structure in Molecules. \emph{Phys. Rev. A} \textbf{1971}, \emph{3},
  565\relax
\mciteBstWouldAddEndPuncttrue
\mciteSetBstMidEndSepPunct{\mcitedefaultmidpunct}
{\mcitedefaultendpunct}{\mcitedefaultseppunct}\relax
\EndOfBibitem
\bibitem[Pettitt(1986)]{Pettitt1986_FirstMcHF}
Pettitt,~B.~A. Hartree-Fock theory of proton states in hydrides. \emph{Chem.
  Phys. Lett.} \textbf{1986}, \emph{130}, 399--402\relax
\mciteBstWouldAddEndPuncttrue
\mciteSetBstMidEndSepPunct{\mcitedefaultmidpunct}
{\mcitedefaultendpunct}{\mcitedefaultseppunct}\relax
\EndOfBibitem
\bibitem[Tachikawa \latin{et~al.}(1998)Tachikawa, Mori, Nakai, and
  Iguchi]{Nakai1998_NOMO-Original}
Tachikawa,~M.; Mori,~K.; Nakai,~H.; Iguchi,~K. {An extension of ab initio
  molecular orbital theory to nuclear motion}. \emph{Chem. Phys. Lett.}
  \textbf{1998}, \emph{290}, 437--442\relax
\mciteBstWouldAddEndPuncttrue
\mciteSetBstMidEndSepPunct{\mcitedefaultmidpunct}
{\mcitedefaultendpunct}{\mcitedefaultseppunct}\relax
\EndOfBibitem
\bibitem[Swalina \latin{et~al.}(2005)Swalina, Pak, and
  Hammes-Schiffer]{Swalina2004-PreBO_MP2}
Swalina,~C.; Pak,~M.~V.; Hammes-Schiffer,~S. {Alternative formulation of
  many-body perturbation theory for electron--proton correlation}. \emph{Chem.
  Phys. Lett.} \textbf{2005}, \emph{404}, 394--399\relax
\mciteBstWouldAddEndPuncttrue
\mciteSetBstMidEndSepPunct{\mcitedefaultmidpunct}
{\mcitedefaultendpunct}{\mcitedefaultseppunct}\relax
\EndOfBibitem
\bibitem[Nakai(2007)]{nakai2007nuclear}
Nakai,~H. {Nuclear orbital plus molecular orbital theory: Simultaneous
  determination of nuclear and electronic wave functions without
  Born--Oppenheimer approximation}. \emph{Int. J. Quantum Chem.} \textbf{2007},
  \emph{107}, 2849--2869\relax
\mciteBstWouldAddEndPuncttrue
\mciteSetBstMidEndSepPunct{\mcitedefaultmidpunct}
{\mcitedefaultendpunct}{\mcitedefaultseppunct}\relax
\EndOfBibitem
\bibitem[Pavo{\v{s}}evi\'{c} \latin{et~al.}(2020)Pavo{\v{s}}evi\'{c}, Rousseau,
  and Hammes-Schiffer]{Pavosevic_OOMP2-Multicomponent}
Pavo{\v{s}}evi\'{c},~F.; Rousseau,~B. J.~G.; Hammes-Schiffer,~S. Multicomponent
  Orbital-Optimized Perturbation Theory Methods: Approaching Coupled Cluster
  Accuracy at Lower Cost. \emph{J. Phys. Chem. Lett.} \textbf{2020}, \emph{11},
  1578--1583\relax
\mciteBstWouldAddEndPuncttrue
\mciteSetBstMidEndSepPunct{\mcitedefaultmidpunct}
{\mcitedefaultendpunct}{\mcitedefaultseppunct}\relax
\EndOfBibitem
\bibitem[Fajen and Brorsen(2021)Fajen, and
  Brorsen]{Fajen2021_MulticomponentMP4}
Fajen,~O.~J.; Brorsen,~K.~R. {Multicomponent MP4 and the inclusion of triple
  excitations in multicomponent many-body methods}. \emph{J. Chem. Phys.}
  \textbf{2021}, \emph{155}, 234108\relax
\mciteBstWouldAddEndPuncttrue
\mciteSetBstMidEndSepPunct{\mcitedefaultmidpunct}
{\mcitedefaultendpunct}{\mcitedefaultseppunct}\relax
\EndOfBibitem
\bibitem[Bochevarov \latin{et~al.}(2004)Bochevarov, Valeev, and
  David~Sherrill]{Valeev2004_ENMO}
Bochevarov,~A.~D.; Valeev,~E.~F.; David~Sherrill,~C. {The electron and nuclear
  orbitals model: current challenges and future prospects}. \emph{Mol. Phys.}
  \textbf{2004}, \emph{102}, 111--123\relax
\mciteBstWouldAddEndPuncttrue
\mciteSetBstMidEndSepPunct{\mcitedefaultmidpunct}
{\mcitedefaultendpunct}{\mcitedefaultseppunct}\relax
\EndOfBibitem
\bibitem[Webb \latin{et~al.}(2002)Webb, Iordanov, and
  Hammes-Schiffer]{Hammes-Schiffer2002}
Webb,~S.~P.; Iordanov,~T.; Hammes-Schiffer,~S. {Multiconfigurational
  nuclear-electronic orbital approach: Incorporation of nuclear quantum effects
  in electronic structure calculations}. \emph{J. Chem. Phys.} \textbf{2002},
  \emph{117}, 4106--4118\relax
\mciteBstWouldAddEndPuncttrue
\mciteSetBstMidEndSepPunct{\mcitedefaultmidpunct}
{\mcitedefaultendpunct}{\mcitedefaultseppunct}\relax
\EndOfBibitem
\bibitem[Cassam-Chena{\"i} \latin{et~al.}(2015)Cassam-Chena{\"i}, Suo, and
  Liu]{Patrick_2015_EN-MFCI}
Cassam-Chena{\"i},~P.; Suo,~B.; Liu,~W. {Decoupling electrons and nuclei
  without the Born-Oppenheimer approximation: The electron-nucleus mean-field
  configuration-interaction method}. \emph{Phys. Rev. A} \textbf{2015},
  \emph{92}, 012502\relax
\mciteBstWouldAddEndPuncttrue
\mciteSetBstMidEndSepPunct{\mcitedefaultmidpunct}
{\mcitedefaultendpunct}{\mcitedefaultseppunct}\relax
\EndOfBibitem
\bibitem[Brorsen(2020)]{Brorsen2020_SelectedCI-PreBO}
Brorsen,~K.~R. {Quantifying multireference character in multicomponent systems
  with heat-bath configuration interaction}. \emph{J. Chem. Theory Comput.}
  \textbf{2020}, \emph{16}, 2379--2388\relax
\mciteBstWouldAddEndPuncttrue
\mciteSetBstMidEndSepPunct{\mcitedefaultmidpunct}
{\mcitedefaultendpunct}{\mcitedefaultseppunct}\relax
\EndOfBibitem
\bibitem[Fajen and Brorsen(2020)Fajen, and
  Brorsen]{Brorsen2020_multicomponentCASSCF}
Fajen,~O.~J.; Brorsen,~K.~R. {Multicomponent CASSCF Revisited: Large Active
  Spaces Are Needed for Qualitatively Accurate Protonic Densities}. \emph{J.
  Chem. Theory Comput.} \textbf{2020}, \emph{17}, 965--974\relax
\mciteBstWouldAddEndPuncttrue
\mciteSetBstMidEndSepPunct{\mcitedefaultmidpunct}
{\mcitedefaultendpunct}{\mcitedefaultseppunct}\relax
\EndOfBibitem
\bibitem[Muolo \latin{et~al.}(2020)Muolo, Baiardi, Feldmann, and
  Reiher]{Muolo2020_Nuclear}
Muolo,~A.; Baiardi,~A.; Feldmann,~R.; Reiher,~M. {Nuclear-electronic
  all-particle density matrix renormalization group}. \emph{J. Chem. Phys.}
  \textbf{2020}, \emph{152}, 204103\relax
\mciteBstWouldAddEndPuncttrue
\mciteSetBstMidEndSepPunct{\mcitedefaultmidpunct}
{\mcitedefaultendpunct}{\mcitedefaultseppunct}\relax
\EndOfBibitem
\bibitem[Feldmann \latin{et~al.}(2022)Feldmann, Muolo, Baiardi, and
  Reiher]{Feldmann2022_QuantumProton}
Feldmann,~R.; Muolo,~A.; Baiardi,~A.; Reiher,~M. {Quantum Proton Effects from
  Density Matrix Renormalization Group Calculations}. \emph{J. Chem. Theory
  Comput.} \textbf{2022}, \emph{18}, 234--250\relax
\mciteBstWouldAddEndPuncttrue
\mciteSetBstMidEndSepPunct{\mcitedefaultmidpunct}
{\mcitedefaultendpunct}{\mcitedefaultseppunct}\relax
\EndOfBibitem
\bibitem[Nakai and Sodeyama(2003)Nakai, and Sodeyama]{Nakai2003}
Nakai,~H.; Sodeyama,~K. {Many-body effects in nonadiabatic molecular theory for
  simultaneous determination of nuclear and electronic wave functions: Ab
  initio NOMO\slash MBPT and CC methods}. \emph{J. Chem. Phys.} \textbf{2003},
  \emph{118}, 1119--1127\relax
\mciteBstWouldAddEndPuncttrue
\mciteSetBstMidEndSepPunct{\mcitedefaultmidpunct}
{\mcitedefaultendpunct}{\mcitedefaultseppunct}\relax
\EndOfBibitem
\bibitem[Ellis \latin{et~al.}(2016)Ellis, Aggarwal, and
  Chakraborty]{Ellis2016_Development}
Ellis,~B.~H.; Aggarwal,~S.; Chakraborty,~A. {Development of the multicomponent
  coupled-cluster theory for investigation of multiexcitonic interactions}.
  \emph{J. Chem. Theory Comput.} \textbf{2016}, \emph{12}, 188--200\relax
\mciteBstWouldAddEndPuncttrue
\mciteSetBstMidEndSepPunct{\mcitedefaultmidpunct}
{\mcitedefaultendpunct}{\mcitedefaultseppunct}\relax
\EndOfBibitem
\bibitem[Pavo{\v{s}}evi{\'c} \latin{et~al.}(2019)Pavo{\v{s}}evi{\'c}, Culpitt,
  and Hammes-Schiffer]{Pavosevic2019_MulticomponentCC}
Pavo{\v{s}}evi{\'c},~F.; Culpitt,~T.; Hammes-Schiffer,~S. {Multicomponent
  Coupled Cluster Singles and Doubles Theory within the Nuclear-Electronic
  Orbital Framework}. \emph{J. Chem. Theory Comput.} \textbf{2019}, \emph{15},
  338--347\relax
\mciteBstWouldAddEndPuncttrue
\mciteSetBstMidEndSepPunct{\mcitedefaultmidpunct}
{\mcitedefaultendpunct}{\mcitedefaultseppunct}\relax
\EndOfBibitem
\bibitem[Pavo{\v{s}}evi{\'c} \latin{et~al.}(2021)Pavo{\v{s}}evi{\'c}, , and
  Hammes-Schiffer]{Pavosevic2021_MulticomponentUCC}
Pavo{\v{s}}evi{\'c},~F.; ; Hammes-Schiffer,~S. {Multicomponent Unitary Coupled
  Cluster and Equation-of-Motion for Quantum Computation}. \emph{J. Chem.
  Theory Comput.} \textbf{2021}, \emph{17}, 3252--3258\relax
\mciteBstWouldAddEndPuncttrue
\mciteSetBstMidEndSepPunct{\mcitedefaultmidpunct}
{\mcitedefaultendpunct}{\mcitedefaultseppunct}\relax
\EndOfBibitem
\bibitem[Pavo{\v{s}}evi{\'c} \latin{et~al.}(2021)Pavo{\v{s}}evi{\'c}, Tao, and
  Hammes-Schiffer]{Hammes-Schiffer2021_MulticomponenDF}
Pavo{\v{s}}evi{\'c},~F.; Tao,~Z.; Hammes-Schiffer,~S. {Multicomponent Coupled
  Cluster Singles and Doubles with Density Fitting: Protonated Water Tetramers
  with Quantized Protons}. \emph{J. Phys. Chem. Lett.} \textbf{2021},
  \emph{12}, 1631--1637\relax
\mciteBstWouldAddEndPuncttrue
\mciteSetBstMidEndSepPunct{\mcitedefaultmidpunct}
{\mcitedefaultendpunct}{\mcitedefaultseppunct}\relax
\EndOfBibitem
\bibitem[Fowler and Brorsen(2022)Fowler, and Brorsen]{Brorsen2022_Triples}
Fowler,~D.; Brorsen,~K.~R. (T) Correction for Multicomponent Coupled-Cluster
  Theory for a Single Quantum Proton. \emph{J. Chem. Theory Comput.}
  \textbf{2022}, \emph{18}, 7298--7305\relax
\mciteBstWouldAddEndPuncttrue
\mciteSetBstMidEndSepPunct{\mcitedefaultmidpunct}
{\mcitedefaultendpunct}{\mcitedefaultseppunct}\relax
\EndOfBibitem
\bibitem[Sirjoosingh \latin{et~al.}(2012)Sirjoosingh, Pak, and
  Hammes-Schiffer]{Hammes-Schiffer2012_MulticomponentDFT}
Sirjoosingh,~A.; Pak,~M.~V.; Hammes-Schiffer,~S. {Multicomponent density
  functional theory study of the interplay between electron-electron and
  electron-proton correlation}. \emph{J. Chem. Phys.} \textbf{2012},
  \emph{136}, 174114\relax
\mciteBstWouldAddEndPuncttrue
\mciteSetBstMidEndSepPunct{\mcitedefaultmidpunct}
{\mcitedefaultendpunct}{\mcitedefaultseppunct}\relax
\EndOfBibitem
\bibitem[Brorsen \latin{et~al.}(2017)Brorsen, Yang, and
  Hammes-Schiffer]{Hammes-Schiffer2017_MulticomponentDFT}
Brorsen,~K.~R.; Yang,~Y.; Hammes-Schiffer,~S. {Multicomponent Density
  Functional Theory: Impact of Nuclear Quantum Effects on Proton Affinities and
  Geometries}. \emph{J. Phys. Chem. Lett.} \textbf{2017}, \emph{8},
  3488--3493\relax
\mciteBstWouldAddEndPuncttrue
\mciteSetBstMidEndSepPunct{\mcitedefaultmidpunct}
{\mcitedefaultendpunct}{\mcitedefaultseppunct}\relax
\EndOfBibitem
\bibitem[Brorsen \latin{et~al.}(2018)Brorsen, Schneider, and
  Hammes-Schiffer]{Brorsen2018_TransferableDFT}
Brorsen,~K.~R.; Schneider,~P.~E.; Hammes-Schiffer,~S. {Alternative forms and
  transferability of electron-proton correlation functionals in
  nuclear-electronic orbital density functional theory}. \emph{J. Chem. Phys.}
  \textbf{2018}, \emph{149}, 044110\relax
\mciteBstWouldAddEndPuncttrue
\mciteSetBstMidEndSepPunct{\mcitedefaultmidpunct}
{\mcitedefaultendpunct}{\mcitedefaultseppunct}\relax
\EndOfBibitem
\bibitem[Tao \latin{et~al.}(2019)Tao, Yang, and
  Hammes-Schiffer]{HammesSchiffer2019_GGA-DFT}
Tao,~Z.; Yang,~Y.; Hammes-Schiffer,~S. {Multicomponent density functional
  theory: Including the density gradient in the electron-proton correlation
  functional for hydrogen and deuterium}. \emph{J. Chem. Phys.} \textbf{2019},
  \emph{151}, 124102\relax
\mciteBstWouldAddEndPuncttrue
\mciteSetBstMidEndSepPunct{\mcitedefaultmidpunct}
{\mcitedefaultendpunct}{\mcitedefaultseppunct}\relax
\EndOfBibitem
\bibitem[Nakai \latin{et~al.}(2005)Nakai, Hoshino, Miyamoto, and
  Hyodo]{Nakai2005_EliminationRotTrans}
Nakai,~H.; Hoshino,~M.; Miyamoto,~K.; Hyodo,~S. Elimination of translational
  and rotational motions in nuclear orbital plus molecular orbital theory.
  \emph{J. Chem. Phys.} \textbf{2005}, \emph{122}, 164101\relax
\mciteBstWouldAddEndPuncttrue
\mciteSetBstMidEndSepPunct{\mcitedefaultmidpunct}
{\mcitedefaultendpunct}{\mcitedefaultseppunct}\relax
\EndOfBibitem
\bibitem[Miyamoto \latin{et~al.}(2006)Miyamoto, Hoshino, and
  Nakai]{Nakai2006_Elimination2}
Miyamoto,~K.; Hoshino,~M.; Nakai,~H. Elimination of translational and
  rotational motions in nuclear orbital plus molecular orbital theory:
  Contribution of the first-order rovibration coupling. \emph{J. Chem. Theory
  Comput.} \textbf{2006}, \emph{2}, 1544--1550\relax
\mciteBstWouldAddEndPuncttrue
\mciteSetBstMidEndSepPunct{\mcitedefaultmidpunct}
{\mcitedefaultendpunct}{\mcitedefaultseppunct}\relax
\EndOfBibitem
\bibitem[Sutcliffe(2005)]{Sutcliffe2005_CommentNakai}
Sutcliffe,~B. Comment on “Elimination of translational and rotational motions
  in nuclear orbital plus molecular orbital theory”[J. Chem. Phys. 122,
  164101 (2005)]. \emph{J. Chem. Phys.} \textbf{2005}, \emph{123}, 164101\relax
\mciteBstWouldAddEndPuncttrue
\mciteSetBstMidEndSepPunct{\mcitedefaultmidpunct}
{\mcitedefaultendpunct}{\mcitedefaultseppunct}\relax
\EndOfBibitem
\bibitem[Blanco and Heller(1983)Blanco, and Heller]{Heller1983_Projection}
Blanco,~M.; Heller,~E.~J. Angular momentum projection operators and molecular
  bound states. \emph{J. Chem. Phys.} \textbf{1983}, \emph{78},
  2504--2517\relax
\mciteBstWouldAddEndPuncttrue
\mciteSetBstMidEndSepPunct{\mcitedefaultmidpunct}
{\mcitedefaultendpunct}{\mcitedefaultseppunct}\relax
\EndOfBibitem
\bibitem[Muolo \latin{et~al.}(2019)Muolo, M{\'a}tyus, and Reiher]{muolo2019}
Muolo,~A.; M{\'a}tyus,~E.; Reiher,~M. {H$_3^+$ as a five-body problem described
  with explicitly correlated Gaussian basis sets}. \emph{J. Chem. Phys.}
  \textbf{2019}, \emph{151}, 154110\relax
\mciteBstWouldAddEndPuncttrue
\mciteSetBstMidEndSepPunct{\mcitedefaultmidpunct}
{\mcitedefaultendpunct}{\mcitedefaultseppunct}\relax
\EndOfBibitem
\bibitem[Ring and Schuck(2004)Ring, and Schuck]{ring_2004nuclear}
Ring,~P.; Schuck,~P. \emph{The nuclear many-body problem}; Springer Science \&
  Business Media, 2004\relax
\mciteBstWouldAddEndPuncttrue
\mciteSetBstMidEndSepPunct{\mcitedefaultmidpunct}
{\mcitedefaultendpunct}{\mcitedefaultseppunct}\relax
\EndOfBibitem
\bibitem[Sheikh and Ring(2000)Sheikh, and Ring]{Sheikh_2000symmetry}
Sheikh,~J.~A.; Ring,~P. Symmetry-projected Hartree--Fock--Bogoliubov equations.
  \emph{Nucl. Phys. A} \textbf{2000}, \emph{665}, 71--91\relax
\mciteBstWouldAddEndPuncttrue
\mciteSetBstMidEndSepPunct{\mcitedefaultmidpunct}
{\mcitedefaultendpunct}{\mcitedefaultseppunct}\relax
\EndOfBibitem
\bibitem[Sheikh \latin{et~al.}(2021)Sheikh, Dobaczewski, Ring, Robledo, and
  Yannouleas]{sheikh2021_SymmetryReview}
Sheikh,~J.~A.; Dobaczewski,~J.; Ring,~P.; Robledo,~L.~M.; Yannouleas,~C.
  Symmetry restoration in mean-field approaches. \emph{J. Phys. G}
  \textbf{2021}, \emph{48}, 123001\relax
\mciteBstWouldAddEndPuncttrue
\mciteSetBstMidEndSepPunct{\mcitedefaultmidpunct}
{\mcitedefaultendpunct}{\mcitedefaultseppunct}\relax
\EndOfBibitem
\bibitem[Bally and Bender(2021)Bally, and Bender]{Bally2021_Projection}
Bally,~B.; Bender,~M. Projection on particle number and angular momentum:
  Example of triaxial Bogoliubov quasiparticle states. \emph{Phys. Rev. C}
  \textbf{2021}, \emph{103}, 024315\relax
\mciteBstWouldAddEndPuncttrue
\mciteSetBstMidEndSepPunct{\mcitedefaultmidpunct}
{\mcitedefaultendpunct}{\mcitedefaultseppunct}\relax
\EndOfBibitem
\bibitem[Morrison and Parker(1987)Morrison, and
  Parker]{Gregory1987_GuideToRotations}
Morrison,~M.~A.; Parker,~G.~A. A guide to rotations in quantum mechanics.
  \emph{Aust. J. Phys.} \textbf{1987}, \emph{40}, 465--498\relax
\mciteBstWouldAddEndPuncttrue
\mciteSetBstMidEndSepPunct{\mcitedefaultmidpunct}
{\mcitedefaultendpunct}{\mcitedefaultseppunct}\relax
\EndOfBibitem
\bibitem[Lestrange \latin{et~al.}(2018)Lestrange, Williams-Young, Petrone,
  Jim{\'e}nez-Hoyos, and Li]{Xiasong2018_EfficientProjected}
Lestrange,~P.~J.; Williams-Young,~D.~B.; Petrone,~A.; Jim{\'e}nez-Hoyos,~C.~A.;
  Li,~X. Efficient implementation of variation after projection generalized
  Hartree--Fock. \emph{J. Chem. Theory Comput.} \textbf{2018}, \emph{14},
  588--596\relax
\mciteBstWouldAddEndPuncttrue
\mciteSetBstMidEndSepPunct{\mcitedefaultmidpunct}
{\mcitedefaultendpunct}{\mcitedefaultseppunct}\relax
\EndOfBibitem
\bibitem[Shimizu and Tsunoda(2023)Shimizu, and
  Tsunoda]{Shimizu2023_SO3Quadrature}
Shimizu,~N.; Tsunoda,~Y. SO(3) quadratures in angular-momentum projection.
  \emph{Comput. Phys. Commun.} \textbf{2023}, \emph{283}, 108583\relax
\mciteBstWouldAddEndPuncttrue
\mciteSetBstMidEndSepPunct{\mcitedefaultmidpunct}
{\mcitedefaultendpunct}{\mcitedefaultseppunct}\relax
\EndOfBibitem
\bibitem[Helgaker \latin{et~al.}(2014)Helgaker, Jorgensen, and
  Olsen]{Helgaker2014_Bible}
Helgaker,~T.; Jorgensen,~P.; Olsen,~J. \emph{Molecular electronic-structure
  theory}; John Wiley \& Sons, 2014\relax
\mciteBstWouldAddEndPuncttrue
\mciteSetBstMidEndSepPunct{\mcitedefaultmidpunct}
{\mcitedefaultendpunct}{\mcitedefaultseppunct}\relax
\EndOfBibitem
\bibitem[Pinchon and Hoggan(2007)Pinchon, and
  Hoggan]{Pinchon2007_RotationSphericalHarmonics}
Pinchon,~D.; Hoggan,~P.~E. Rotation matrices for real spherical harmonics:
  general rotations of atomic orbitals in space-fixed axes. \emph{J. Phys. A:
  Math. Theor} \textbf{2007}, \emph{40}, 1597\relax
\mciteBstWouldAddEndPuncttrue
\mciteSetBstMidEndSepPunct{\mcitedefaultmidpunct}
{\mcitedefaultendpunct}{\mcitedefaultseppunct}\relax
\EndOfBibitem
\bibitem[Thom and Head-Gordon(2009)Thom, and
  Head-Gordon]{Thom2009_Nonorthogonal}
Thom,~A.~J.; Head-Gordon,~M. Hartree--Fock solutions as a quasidiabatic basis
  for nonorthogonal configuration interaction. \emph{J. Chem. Phys.}
  \textbf{2009}, \emph{131}, 124113\relax
\mciteBstWouldAddEndPuncttrue
\mciteSetBstMidEndSepPunct{\mcitedefaultmidpunct}
{\mcitedefaultendpunct}{\mcitedefaultseppunct}\relax
\EndOfBibitem
\bibitem[Jim{\'e}nez-Hoyos \latin{et~al.}(2012)Jim{\'e}nez-Hoyos, Henderson,
  Tsuchimochi, and Scuseria]{Jimenez2012_projected}
Jim{\'e}nez-Hoyos,~C.~A.; Henderson,~T.~M.; Tsuchimochi,~T.; Scuseria,~G.~E.
  Projected hartree--fock theory. \emph{J. Chem. Phys.} \textbf{2012},
  \emph{136}, 164109\relax
\mciteBstWouldAddEndPuncttrue
\mciteSetBstMidEndSepPunct{\mcitedefaultmidpunct}
{\mcitedefaultendpunct}{\mcitedefaultseppunct}\relax
\EndOfBibitem
\bibitem[L\"owdin(1955)]{Loewdin_NO_rules}
L\"owdin,~P.-O. Quantum Theory of Many-Particle Systems. II. Study of the
  Ordinary Hartree-Fock Approximation. \emph{Phys. Rev.} \textbf{1955},
  \emph{97}, 1490--1508\relax
\mciteBstWouldAddEndPuncttrue
\mciteSetBstMidEndSepPunct{\mcitedefaultmidpunct}
{\mcitedefaultendpunct}{\mcitedefaultseppunct}\relax
\EndOfBibitem
\bibitem[Feldmann \latin{et~al.}(2023)Feldmann, Baiardi, and
  Reiher]{Feldmann2023_Second}
Feldmann,~R.; Baiardi,~A.; Reiher,~M. Second-Order Self-Consistent Field
  Algorithms: From Classical to Quantum Nuclei. \emph{J. Chem. Theory Comput.}
  \textbf{2023}, \emph{19}, 856--873\relax
\mciteBstWouldAddEndPuncttrue
\mciteSetBstMidEndSepPunct{\mcitedefaultmidpunct}
{\mcitedefaultendpunct}{\mcitedefaultseppunct}\relax
\EndOfBibitem
\bibitem[Alml{\"o}f \latin{et~al.}(1982)Alml{\"o}f, F{\ae}gri~Jr, and
  Korsell]{Almlof1982_SAD}
Alml{\"o}f,~J.; F{\ae}gri~Jr,~K.; Korsell,~K. {Principles for a direct SCF
  approach to LCAO--MO \textit{ab-initio} calculations}. \emph{J. Comput.
  Chem.} \textbf{1982}, \emph{3}, 385--399\relax
\mciteBstWouldAddEndPuncttrue
\mciteSetBstMidEndSepPunct{\mcitedefaultmidpunct}
{\mcitedefaultendpunct}{\mcitedefaultseppunct}\relax
\EndOfBibitem
\bibitem[Dunning~Jr.(1989)]{Dunning1989_gaussian}
Dunning~Jr.,~T.~H. {Gaussian basis sets for use in correlated molecular
  calculations. I. The atoms boron through neon and hydrogen}. \emph{J. Chem.
  Phys.} \textbf{1989}, \emph{90}, 1007--1023\relax
\mciteBstWouldAddEndPuncttrue
\mciteSetBstMidEndSepPunct{\mcitedefaultmidpunct}
{\mcitedefaultendpunct}{\mcitedefaultseppunct}\relax
\EndOfBibitem
\bibitem[Samsonova \latin{et~al.}(2023)Samsonova, Tucker, Alaal, and
  Brorsen]{Brorsen2023_BasisSets}
Samsonova,~I.; Tucker,~G.~B.; Alaal,~N.; Brorsen,~K.~R. Hydrogen-Atom
  Electronic Basis Sets for Multicomponent Quantum Chemistry. \emph{ACS omega}
  \textbf{2023}, \emph{8}, 5033--5041\relax
\mciteBstWouldAddEndPuncttrue
\mciteSetBstMidEndSepPunct{\mcitedefaultmidpunct}
{\mcitedefaultendpunct}{\mcitedefaultseppunct}\relax
\EndOfBibitem
\bibitem[Yu \latin{et~al.}(2020)Yu, Pavo{\v{s}}evi{\'c}, and
  Hammes-Schiffer]{Hammes-Schiffer2020_NuclearBasis}
Yu,~Q.; Pavo{\v{s}}evi{\'c},~F.; Hammes-Schiffer,~S. {Development of nuclear
  basis sets for multicomponent quantum chemistry methods}. \emph{J. Chem.
  Phys.} \textbf{2020}, \emph{152}, 244123\relax
\mciteBstWouldAddEndPuncttrue
\mciteSetBstMidEndSepPunct{\mcitedefaultmidpunct}
{\mcitedefaultendpunct}{\mcitedefaultseppunct}\relax
\EndOfBibitem
\bibitem[Feldmann \latin{et~al.}(2023)Feldmann, Baiardi, Bosia, and
  Reiher]{kiwi}
Feldmann,~R.; Baiardi,~A.; Bosia,~F.; Reiher,~M. qcscine/kiwi: Release 1.0.0.
  2023; \url{https://doi.org/10.5281/zenodo.7569669}\relax
\mciteBstWouldAddEndPuncttrue
\mciteSetBstMidEndSepPunct{\mcitedefaultmidpunct}
{\mcitedefaultendpunct}{\mcitedefaultseppunct}\relax
\EndOfBibitem
\bibitem[Feldmann \latin{et~al.}(2023)Feldmann, Baiardi, Bosia, and
  Reiher]{integralevaluator}
Feldmann,~R.; Baiardi,~A.; Bosia,~F.; Reiher,~M. qcscine/integralevaluator:
  Release 1.0.0. 2023; \url{https://doi.org/10.5281/zenodo.7569657}\relax
\mciteBstWouldAddEndPuncttrue
\mciteSetBstMidEndSepPunct{\mcitedefaultmidpunct}
{\mcitedefaultendpunct}{\mcitedefaultseppunct}\relax
\EndOfBibitem
\bibitem[Valeev(2022)]{Libint2}
Valeev,~E.~F. Libint: A library for the evaluation of molecular integrals of
  many-body operators over Gaussian functions. http://libint.valeyev.net/,
  2022; version 2.7.2\relax
\mciteBstWouldAddEndPuncttrue
\mciteSetBstMidEndSepPunct{\mcitedefaultmidpunct}
{\mcitedefaultendpunct}{\mcitedefaultseppunct}\relax
\EndOfBibitem
\bibitem[Lindsay and McCall(2001)Lindsay, and
  McCall]{Lindsay2001_ComprehensiveH3plus}
Lindsay,~C.~M.; McCall,~B.~J. Comprehensive evaluation and compilation of
  H\textsubscript{3}\textsuperscript{+} spectroscopy. \emph{J. Mol. Spectrosc.}
  \textbf{2001}, \emph{210}, 60--83\relax
\mciteBstWouldAddEndPuncttrue
\mciteSetBstMidEndSepPunct{\mcitedefaultmidpunct}
{\mcitedefaultendpunct}{\mcitedefaultseppunct}\relax
\EndOfBibitem
\bibitem[Sutcliffe(2021)]{Sutcliffe2021_Respect}
Sutcliffe,~B. Treating nuclei in molecules with quantum mechanical respect.
  \emph{Theor. Chem. Acc.} \textbf{2021}, \emph{140}, 23\relax
\mciteBstWouldAddEndPuncttrue
\mciteSetBstMidEndSepPunct{\mcitedefaultmidpunct}
{\mcitedefaultendpunct}{\mcitedefaultseppunct}\relax
\EndOfBibitem
\end{mcitethebibliography}
\providecommand{\latin}[1]{#1}
\makeatletter
\providecommand{\doi}
  {\begingroup\let\do\@makeother\dospecials
  \catcode`\{=1 \catcode`\}=2 \doi@aux}
\providecommand{\doi@aux}[1]{\endgroup\texttt{#1}}
\makeatother
\providecommand*\mcitethebibliography{\thebibliography}
\csname @ifundefined\endcsname{endmcitethebibliography}
  {\let\endmcitethebibliography\endthebibliography}{}

\end{document}